\documentclass[twocolumn,showpacs,preprintnumbers,
               superscriptaddress]{revtex4}
\usepackage{graphicx,latexsym}

\newcommand{\Tr}{\mbox{Tr}}
\renewcommand{\Re}{\mbox{Re}}

\begin{document}

\preprint{YITP-03-44}

\title{Negative-parity baryon spectra in quenched anisotropic lattice QCD}

\author{Y.~Nemoto\footnote{
Present address: RIKEN BNL Research Center, BNL, Upton, NY 11973}}
\affiliation{Yukawa Institute for Theoretical Physics,
             Kyoto University, Kyoto 606-8502, Japan}

\author{N.~Nakajima}
\affiliation{Center of Medical Information Science,
             Kochi Medical School, Kochi, 783-8505 Japan}

\author{H.~Matsufuru}
\affiliation{Yukawa Institute for Theoretical Physics,
             Kyoto University, Kyoto 606-8502, Japan}

\author{H.~Suganuma}
\affiliation{Faculty of Science, Tokyo Institute of Technology,
             Tokyo 152-8551, Japan}


\pacs{12.38.Gc, 14.20.Gk, 14.20.Jn}

\begin{abstract}
We investigate the negative-parity baryon spectra in quenched lattice QCD.
We employ the anisotropic lattice with standard Wilson gauge and
$O(a)$ improved Wilson quark actions at three values of lattice
spacings with renormalized anisotropy $\xi=a_{\sigma}/a_{\tau}=4$,
where $a_{\sigma}$ and $a_{\tau}$ are spatial and temporal lattice
spacings, respectively.
The negative-parity baryons are measured with the parity projection. 
In particular, we pay much attention to the lowest $SU(3)$
flavor-singlet negative-parity baryon, 
which is assigned as the $\Lambda(1405)$ in the quark model.
For the flavor octet and decuplet negative-parity baryons, 
the calculated masses are close to the experimental values of
corresponding lowest-lying negative-parity baryons.
In contrast, the flavor-singlet baryon is found to be about 1.7 GeV, 
which is much heavier than the $\Lambda(1405)$. 
Therefore, it is difficult to identify the $\Lambda(1405)$ to be the 
flavor-singlet three-quark state, 
which seems to support an interesting picture of the penta-quark 
($uds q\bar q$) state or the $N\bar K$ molecule for the $\Lambda(1405)$.
\end{abstract}

\maketitle

\section{Introduction}
 \label{sec:Introduction}

The lattice QCD simulation has become a powerful method to investigate 
hadron properties directly based on QCD.
The spectroscopy of lowest-lying hadrons in the quenched
approximation, i.e., without dynamical quark-loop effects, 
has been almost established, and reproduces their experimental values 
within 10 \% deviations \cite{CPPACS}.
Extensive simulations including dynamical quarks are in progress and
would give us detailed understanding of 
the spectra of these ground-state hadrons \cite{FullQCD}.
In contrast, several lattice studies on the excited-state hadrons have been 
started \cite{NMNS01,LL99,SBO01,Goc01,Mel02,Lee02,UK02} 
very recently, and their calculations are far from established even at
the quenched level.
In this paper, using anisotropic lattice QCD, we investigate the
low-lying negative-parity baryon spectra, 
particularly paying attention to the flavor-singlet baryon. 
Another purpose of this paper is to examine the correspondence between 
the flavor-singlet three-quark (3Q) state and the $\Lambda(1405)$.

In the context of the flavor-singlet baryon, the $\Lambda(1405)$ is
known to be one of the most mysterious hadrons. 
The $\Lambda(1405)$ is the lightest negative-parity baryon, although
it contains strangeness.
Moreover, there are two physical interpretations on the $\Lambda(1405)$. 
From the viewpoint of the quark model, the $\Lambda(1405)$ is
described as the flavor-singlet 3Q system.
As another interpretation, the $\Lambda(1405)$ is an interesting
candidate of the hadronic molecule such as the 
$N \bar K$ bound state with a large binding energy of about 30 MeV.
We aim to clarify whether the $\Lambda(1405)$ can be explained as
a flavor-singlet baryon in quenched lattice QCD.

Historically, excited-state baryons have been so far mainly
investigated within the framework of the non-relativistic quark model,
in which baryons can be classified in terms of the spin-flavor
$SU(6)$ symmetry. 
While the ground-state baryons have completely symmetric spin-flavor
wave functions and form the 56-dimensional representation,
the low-lying negative-parity baryons are parts of the $L=1$ orbitally
excited states and belong to the $SU(6)$ 70-dimensional representation.
We summarize the classification of the $SU(6)$ symmetry and its
assignment to experimentally observed baryons in Table \ref{tab:su6baryons}.
Both the non-relativistic \cite{IK79} and the semi-relativistic
\cite{CI86} quark models reproduce the negative-parity baryon spectra fairly
well in the octet and the decuplet sectors \cite{CR00}.
Such success of the quark model implies that the constituent quark
picture holds well and the gluonic excitation modes play a less
important role. 
The potential form is being clarified with the recent lattice QCD
calculations of the static 3Q potential \cite{TMNS01,IBSS02}.
Furthermore, the large gluonic excitation energy \cite{TS02} obtained
with lattice QCD explains the reason why the quark  
potential model without such gluonic excitations well describes
the hadron spectra.

Among the low-lying negative-parity baryons, the $\Lambda(1405)$ is
an exception of such success of the quark potential model. 
In fact, the $\Lambda(1405)$ is much lighter than the lowest-lying
non-strange negative-parity baryons, 
$N(1520)$ with $J^P=3/2^-$ and $N(1535)$ with $J^P=1/2^-$.
There are two physical interpretations proposed for the
$\Lambda(1405)$: an $SU(3)$ flavor-singlet 3Q state, 
and an $N\bar{K}$ bound state, i.e., a penta-quark (5Q) system.
The simple quark model is based on the former picture, and it predicts
that the $\Lambda(1405)$ and the $\Lambda(1520)$
with $J^P=3/2^-$ are nearly degenerate \cite{IK79,CI86}.
In this picture, the large mass difference between them is explained
to originate from a large $LS$-force \cite{PDG02}, 
but such a strong $LS$ splitting is not observed in other baryon
spectra, and therefore 
it seems difficult to reproduce the mass of the $\Lambda(1405)$ within
the simple quark model.

Another interesting interpretation for the $\Lambda(1405)$ is 
the penta-quark (5Q) system or the $N\bar{K}$ bound state as a
hadronic molecule~\cite{PDG02, Ch98}.
Note here that the $\Lambda(1405)$ lies about 30 MeV below the
$N\bar{K}$ threshold, 
and this binding energy of about 30 MeV is rather large in comparison
with about 2.2 MeV, that of the deuteron.
If this picture holds true, the large binding energy between
$N$ and $\bar K$ results in a significant role of the attractive
effect for the $\bar K$ put inside nuclei or nuclear matter.
In this way, the study of such an exotic and strange baryon,
the $\Lambda(1405)$, is important also for understanding of the 
manifestation of strangeness in the hyper-nuclei and the neutron stars.

The 3Q and the 5Q states, however, would mix in the real world.
Therefore, a more realistic question would be as follows.
Which is the dominant component of the $\Lambda(1405)$,
the 3Q state or the 5Q state?
We try to answer this question using lattice QCD simulations.
In lattice QCD simulations, even if one chooses the operator
as the 3Q or the 5Q state, it generally overlaps with both the states
through the quark and anti-quark pair creation.
In the quenched simulation, however, owing to the absence of dynamical
quark-loop effects,
such a mixing between 3Q and 5Q states are rather suppressed, which
would enable us to investigate   
the properties of genuine 3Q and 5Q states in a separate manner. 

In this paper, we focus on the 3Q state and investigate whether
this picture can explain the mass of the $\Lambda(1405)$.
Apparent discrepancy with the experimentally observed mass implies 
that the penta-quark state gives significant contribution to the physical
$\Lambda(1405)$ state.
In practice, lattice QCD results suffer from various systematic errors.
It is therefore essential to compare the mass of the flavor-singlet
3Q state with other negative-parity baryon masses
as well as with lowest-lying baryon masses.

It is also important to understand the gross structure of low-lying
negative-parity baryon spectrum in relation to spontaneous
chiral-symmetry breaking in QCD.
If the chiral symmetry is restored, such as at high temperature
and/or density, the masses of a baryon and its parity partner
should be degenerate.
Spontaneous chiral-symmetry breaking causes mass 
splitting between positive- and negative- parity baryons.
It is important to study nonperturbatively negative-parity baryons 
in terms of the parity partners of the positive-parity ones.

Since the negative-parity baryons have relatively large masses, 
their correlators rapidly decrease in the Euclidean temporal direction.
As a technical improvement, we adopt an anisotropic lattice 
where the temporal lattice spacing $a_{\tau}$ is finer than the
spatial one, $a_{\sigma}$ \cite{Kar82}.
With the high resolution in the temporal direction, we can follow
the change of the correlator in detail and specify the relevant
region for extraction of the mass.
Thus efficient measurements would be possible.
This approach is efficient also for other correlators of heavy
particles, such as the glueballs~\cite{ISM02}. 
(The anisotropic lattice is extremely powerful for the study of the
finite temperature QCD~\cite{ISM02, Ume01,TAROFT},
where the temporal distance is severely limited in the imaginary-time
formalism.)

In this study, we adopt the standard Wilson plaquette gauge action
and $O(a)$ improved Wilson quark action, for which the sizes of
errors are rather well evaluated \cite{Kla98,Aniso01b}.
The simulations are performed on the quenched anisotropic lattices with
renormalized anisotropy $\xi=a_{\sigma}/a_{\tau}=4$ at three
lattice spacings in the range of $a_{\sigma}^{-1}\simeq$1--2 GeV.
For these lattices, the quark parameters were tuned and the light
hadron spectrum was calculated 
in order to estimate the effects of uncertainties due to anisotropy on
the spectrum~\cite{Aniso01b}.

This paper is organized as follows.
In Section~\ref{sec:aniso_lattice}, we summarize the anisotropic
lattice actions used in this study.
We show the numerical results on the negative-parity baryon in
Section~\ref{sec:simulation},
and discuss their physical consequences in Section~\ref{sec:discussion}.
The last section is dedicated to the conclusion and perspective
for further studies.

\begin{table*}
\caption{Quark model assignments for experimentally observed
baryons in terms of the spin-flavor SU(6) basis~\cite{PDG02}.
}
\label{tab:su6baryons}
\begin{ruledtabular}
\begin{tabular}{cccccccc}
$SU(6)$ rep. & $SU(3)_f$ rep.  & $J^P$ & $S=0$ & $S=-1,I=0$ & $S=-1,I=1$ & 
$S=-2$ & $S=-3$  \\
\hline
56 ($L=0$) & ${}^2 8$  & $\frac{1}{2}^+$ & $N(939)$  & $\Lambda(1116)$ &
$\Sigma(1193)$ & $\Xi(1318)$ &  \\
           & ${}^4 10$ & $\frac{3}{2}^+$ & $\Delta(1232)$ & &
$\Sigma(1385)$ & $\Xi(1530)$ & $\Omega(1672)$ \\
\hline 
70 ($L=1$) & ${}^2 8$  & $\frac{1}{2}^-$ & $N(1535)$ & $\Lambda(1670)$ &
$\Sigma(1620)$ & $\Xi(?)$  & \\
           &           & $\frac{3}{2}^-$ & $N(1520)$ & $\Lambda(1690)$ &
$\Sigma(1670)$ & $\Xi(1820)$ & \\
           & ${}^4 8$  & $\frac{1}{2}^-$ & $N(1650)$ & $\Lambda(1800)$ &
$\Sigma(1750)$ & $\Xi(?)$ & \\
           &           & $\frac{3}{2}^-$ & $N(1700)$ & $\Lambda(?)$ &
$\Sigma(?)$ & $\Xi(?)$ & \\
           &           & $\frac{5}{2}^-$ & $N(1675)$ & $\Lambda(1830)$ &
$\Sigma(1775)$ & $\Xi(?)$ & \\
           & ${}^2 10$ & $\frac{1}{2}^-$ & $\Delta(1620)$ & & $\Sigma(?)$ 
& $\Xi(?)$ & $\Omega(?)$ \\
           &           & $\frac{3}{2}^-$ & $\Delta(1700)$ & & $\Sigma(?)$
& $\Xi(?)$ & $\Omega(?)$ \\
           & ${}^2 1$  & $\frac{1}{2}^-$ & & $\Lambda(1405)$ & & & \\
           &           & $\frac{3}{2}^-$ & & $\Lambda(1520)$ & & &
\end{tabular}
\end{ruledtabular}
\end{table*}

\section{Anisotropic lattice}
 \label{sec:aniso_lattice}

We employ the standard Wilson plaquette gauge action
and the $O(a)$ improved Wilson quark action on anisotropic lattices.
We briefly summarize the anisotropic lattice action.
The gauge field action takes the from
\begin{eqnarray}
S_G &=& \beta \sum_x \left\{
  \sum_{i>j=1}^{3}  \frac{1}{\gamma_G}
          \left[ 1 - \frac{1}{3} \Re \Tr U_{ij}(x)  \right]
  \right.
 \nonumber \\
 & & \hspace{1.2cm}  \left.
  + \sum_{i=1}^{3} \gamma_G
          \left[ 1 - \frac{1}{3} \Re \Tr U_{i4}(x)  \right]
 \right\}
 \label{eq:gauge_action}
\end{eqnarray}
with $\beta \equiv 2N_c/g^2$. Here, $U_{\mu\nu}$ denotes the parallel
transport around a plaquette in $\mu$-$\nu$ plane,
\begin{equation}
U_{\mu\nu}(x) = U_{\mu}(x) U_{\nu}(x+\hat{\mu})
                U_{\mu}^{\dag}(x+\hat{\nu}) U_{\nu}^{\dag}(x).
\end{equation}
The gluon field is represented with the link-variable as
$U_{\mu}\simeq \exp(-iga_{\mu}A_{\mu})$.
The bare anisotropy $\gamma_G$ coincides with the renormalized
anisotropy $\xi=a_{\sigma}/a_{\tau}$ at the tree level.

Note here that the bare anisotropy is no longer the same as
$\xi$ due to the quantum effect, and one needs to measure $\xi$
through some physical observables for each input value of $\gamma_G$.
Although $\xi$ is in general a function of gauge and quark parameters,
($\beta,\gamma_G$) and ($\kappa, \gamma_F$) respectively,
on a quenched lattice, 
the calibrations of the gauge and quark actions can be performed separately.
For the gauge action of the form (\ref{eq:gauge_action}),
Klassen nonperturbatively obtained an expression of $\gamma_G$
in terms of $\beta$ and $\xi$ with the accuracy better than 1\% 
using the Wilson loops~\cite{Kla98}.
We adopt the same lattice actions as those in Ref.\cite{Aniso01b}
which made use of the Klassen's result for the gauge action
and also performed sufficient analysis for the quark actions
as mentioned below.

For the Wilson type quark action, $O(a)$ improvement
is significant in quantitative computation of hadron spectrum.
Among several types of the anisotropic lattice quark action,
we use the form proposed in Refs.~\cite{Ume01,Aniso01a,Aniso01b}.
As a merit of  this form, the calibration with a good precision
is rather easy in the light quark mass region, since the quark mass
dependence is expected to be small there, 
as was numerically shown in Ref.~\cite{Aniso01b}.

The quark action is written as
\begin{equation}
S_F = \sum_{x,y} \bar{\psi}(x) K(x,y) \psi(y),
\vspace{-0.5cm}
\end{equation}
\begin{eqnarray}
K(x,y)  &=&  \delta_{x,y} \nonumber \\
 & & \hspace{-1.6cm}
 - \kappa_{\tau} \left\{
     (1\!-\!\gamma_4) U_4(x) \delta_{x+\hat{4},y}
   + (1\!+\!\gamma_4) U_4^{\dag}(x-\hat{4}) \delta_{x\!-\!\hat{4},y}
  \right\}
 \nonumber \\
 & & \hspace{-1.6cm}
  - \kappa_{\sigma} \sum_i \left\{
     (r\!-\!\gamma_i) U_i(x) \delta_{x+\hat{i},y}
   + (r\!+\!\gamma_i) U_i^{\dag}(x-\hat{i}) \delta_{x\!-\!\hat{i},y}
  \right\}
 \nonumber \\
 & & \hspace{-1.6cm}
  - \kappa_{\sigma} c_E \sum_i \sigma_{i4} F_{i4} \delta_{x,y}
  - r \kappa_{\sigma} c_B \sum_{i>j} \sigma_{ij} F_{ij} \delta_{x,y},
 \label{eq:action} 
\end{eqnarray}
where $\psi$ denotes the anticommuting quark field, 
$\kappa_{\sigma}$ and $\kappa_{\tau}$ 
the spatial and temporal hopping parameters, respectively, $r$ the spatial
Wilson parameter and $c_E$, $c_B$ the clover coefficients.
The field strength $F_{\mu\nu}$ is defined with the standard
clover-leaf-type construction.
In principle, for a given $\kappa_{\sigma}$, the 
four parameters $\kappa_{\sigma}/\kappa_{\tau}$, $r$, $c_E$ and $c_B$
should be tuned so that Lorentz symmetry is satisfied up to 
discretization errors of $O(a^2)$.
Following Refs. \cite{Aniso01b,Ume01,Aniso01a},
we set the spatial Wilson parameter as $r=1/\xi$ and
the clover coefficients as the tadpole-improved tree-level values, namely,
\begin{equation}
 r = 1/\xi, \hspace{0.5cm}
 c_E= 1/u_{\sigma} u_{\tau}^2, \hspace{0.5cm}
 c_B = 1/u_{\sigma}^3.
\label{eq:cecb}
\end{equation}
To reduce large contribution from the tadpole diagram,
the tadpole improvement \cite{LM93} is applied
by rescaling the link variables as
$U_i(x) \rightarrow U_i(x)/u_{\sigma}$ and  $U_4(x) \rightarrow
U_4(x)/u_{\tau}$, with the mean-field values of the spatial 
and temporal link variables, $u_{\sigma}$ and $u_{\tau}$,
respectively.
This is equivalent to redefining the
hopping parameters with the tadpole-improved ones (with tilde)
through $\kappa_{\sigma} = \tilde{\kappa}_{\sigma}/u_{\sigma}$
and $\kappa_{\tau} = \tilde{\kappa}_{\tau}/u_{\tau}$.
We define the anisotropy parameter $\gamma_F$ as
$\gamma_F \equiv \tilde{\kappa}_{\tau}/\tilde{\kappa}_{\sigma}$.
This parameter in the action is to be tuned nonperturbatively 
in the numerical simulation.
It is convenient to define $\kappa$ as 
\begin{eqnarray}
\frac{1}{\kappa} \equiv \frac{1}{\tilde{\kappa}_{\sigma}}
     - 2(\gamma_F+3r-4) = 2(m_0 \gamma_F + 4),
 \label{eq:kappa}
\end{eqnarray}
where $m_{0}$ is the bare quark mass in temporal lattice units.
This $\kappa$ plays the same role as in the case of isotropic
lattice, and is convenient to parameterize the quark mass together with
the bare anisotropy $\gamma_F$.

The above action is constructed following the Fermilab approach
\cite{EKM97}, which proposes to tune the bare anisotropy
parameter so that the rest mass and the kinetic mass equal each other.
In practice, hadronic states are convenient to carry out this program.
In Ref.~\cite{Aniso01b}, the bare anisotropy was tuned
nonperturbatively using the relativistic dispersion relation
of the pseudoscalar and vector mesons.
The main result of Ref. \cite{Aniso01b} is as follows:
They tuned $\gamma_F$ in the quark mass range from the strange
to charm quark masses with the accuracy better than 1\%,
and found that the tuned bare anisotropy, $\gamma_F^*$, is
well fitted to the linear form in $m_q^2$, where
$m_q = (\kappa^{-1}-\kappa_c^{-1})/2\xi$
is naively defined quark mass.
Then $\gamma_F^*$ in the massless limit was obtained within 
2\% error, i.e., 1\% as a statistical error and 1\% as a systematic
error in the form of fit in terms of $m_q$.
Then, they computed the light hadron spectrum using the
value of $\gamma_F^*$ at the chiral limit, and observed
the effect of uncertainty in $\gamma_F^*$ on the spectrum
for physical quark masses of the 1\%-level.
In the present study, we also treat the same quark mass region and
therefore adopt the value of $\gamma_F^*$ in the chiral limit.
The precision of 2\% error in $\gamma_F^*$ is sufficient for
the present purpose.

As was pointed out in Refs.~\cite{Ume01,Aniso01b},
with the choice $r=1/\xi$, the action (\ref{eq:action}) leads to a
smaller spatial Wilson term for a larger anisotropy $\xi$.
Since the negative-parity baryons measured in this paper are the 
lowest state of the parity projected baryon correlators,
the statements in Ref.~\cite{Aniso01b} for light hadrons
also hold in our calculation.
Even for the coarsest lattice in our calculation, 
the cutoff $a_{\tau}^{-1}\simeq 4.0$ GeV seems sufficiently large to avoid 
the artificial excitation due to the doublers in the ground-state signals. 
At least, the finest lattice with $a_{\tau}^{-1}\simeq 8$ GeV would
be sufficiently large to avoid the doubler effect.

\section{Numerical Simulation}
 \label{sec:simulation}

\subsection{Lattice setup}

\begin{table*}
\caption{
Lattice parameters in the gluon sector.
The scale $a_{\sigma}^{-1}(m_{K^*})$ is determined from the $K^*$ meson 
mass.
The mean-field values are defined in the Landau gauge.
The statistical uncertainty of  $u_{\tau}$ is less than the last digit.
The details for these parameters are described in \cite{Aniso01b}.}
\begin{ruledtabular}
\begin{tabular}{cccccc}
$\beta$ & $\gamma_G$ & size &  $u_{\sigma}$ & $u_{\tau}$ 
& $a_{\sigma}^{-1}(m_{K^*})$ [GeV] \\
\hline
5.75 & 3.072~ & $12^3\times ~96$ & 0.7620(2) & 0.9871 &
  1.034( 6)\\
5.95 & 3.1586 & $16^3\times 128$ & 0.7917(1) & 0.9891 &
  1.499( 9)\\
6.10 & 3.2108 & $20^3\times 160$ & 0.8059(1) & 0.9901 &
  1.871(14)\\
\end{tabular}
\end{ruledtabular}
\label{tab:parameters}
\end{table*}

\begin{table*}
\caption{Lattice parameters in the quark sector.
The values of $\gamma_F$ and $\kappa_c$ are taken from Ref. \cite{Aniso01b}.}
\begin{ruledtabular}
\begin{tabular}{ccccc}
$\beta$ & $\gamma_F$ & $\kappa_c$ &
  $N_{\mbox{conf}}$  & values of $\kappa$ \\
\hline
5.75& 3.909 & 0.12640(5) & 400 & 0.1240, 0.1230, 0.1220, 0.1210 \\
5.95& 4.016 & 0.12592(6) & 400 & 0.1245, 0.1240, 0.1235, 0.1230 \\
6.10& 4.034 & 0.12558(4) & 400 & 0.1245, 0.1240, 0.1235, 0.1230 \\
\end{tabular}
\end{ruledtabular}
\label{tab:parameters2}
\end{table*}

The numerical simulation is performed on the same lattices as
in Ref.~\cite{Aniso01b}.
Here, we briefly summarize the fundamental parameters and
physical quantities.
We use the three anisotropic lattices with the 
renormalized anisotropy $\xi=4$,
at the quenched level.
The statistical uncertainties are, otherwise noted, estimated by the 
Jackknife method with appropriate binning.

The lattice sizes and parameters in generating the gauge field
configurations are listed in Table~\ref{tab:parameters}.
The spatial lattice scales $a_\sigma^{-1}$ roughly cover 1--2 GeV.
The values of bare gluonic anisotropy $\gamma_G$ are chosen
according to the result by Klassen \cite{Kla98}.
The uncertainty of his expression is of the order of 1\%,
and in the following analysis we do not include this uncertainty
in the quoted statistical errors.
The gauge configurations are separated by 2000 (1000) pseudo-heat-bath sweeps,
after 20000 (10000) thermalization sweeps at $\beta$=5.95 and 6.10
(5.75).
The configurations are fixed to the Coulomb gauge,
which is convenient in applying the smearing of hadron operators.

The mean-field values of link variables are determined on the
smaller lattices with half size in temporal extent 
and otherwise with the same parameters for $\beta=5.75$
and $5.95$,
while at $\beta=6.10$ the lattice size is $16^3\times 64$.
The mean-field values, $u_{\sigma}$ and $u_{\tau}$, are obtained
as the averages of the link variables in the Landau gauge,
where the mean-field values are self-consistently used
in the fixing condition \cite{Ume01}.
In a study of hadron spectrum, it is convenient to define the
lattice scales through a hadronic quantity.
We determine $a_{\sigma}^{-1}$
through the $K^*$ meson mass, $m_{K^*}=893.9$ MeV(isospin averaged).
The procedure is the same as in Ref.~\cite{Aniso01b}, while with
larger statistics.
The result is quoted in Table~\ref{tab:parameters} as
$a_\sigma^{-1}(m_{K^*})$.

The quark parameters are listed in Table~\ref{tab:parameters2}.
These hopping parameters roughly cover the quark masses
$m_q\simeq m_s$--$2m_s$.
The numbers of configurations are larger than those of
the hadronic spectroscopy in Ref.~\cite{Aniso01b}.
As already noted in the previous section, the values of $\gamma_F$
and $\kappa_c$ are taken from the result of Ref.~\cite{Aniso01b}.
Although the uncertainty of 2\% level is associated with the values
of $\gamma_F$, the quoted errors of hadron masses in the following
analysis do not include this uncertainty.
According to Ref.~\cite{Aniso01b}, this uncertainty
in the physical masses of vector mesons and positive-parity baryons
are at most of order of 1\%.
For the negative-parity baryon masses, the effect is expected to be
similar amount and not significant compared with the present
level of statistical error.
Therefore, this effect can be negligible in our calculation.

As described later, we first extrapolate the vector meson mass
linearly in the pseudoscalar meson mass squared to the point at which
the ratio of these meson masses are equal to the physical value,
$m_{K^*}/m_K$.
At this point, aforementioned lattice scale is determined.
The physical ($u$, $d$) and $s$ quark masses are determined through
the $\pi$ and $K$ meson masses, $m_{\pi}^{\pm}=139.6$ MeV and
$m_K=495.7$ MeV (isospin averaged), respectively.

\subsection{Baryon correlators}

We measure the correlators in pseudoscalar and vector meson channels
and octet ($\Sigma$ and $\Lambda$ types), decuplet and
singlet channels of SU(3) flavor representation of baryons.
As listed in Table~\ref{tab:operator},
we use the standard meson and baryon operators which have the same
quantum numbers as the corresponding baryons and survive
in the non-relativistic limit.
It is known that there are mainly two ways to choose the baryon
operator.
One is the operator taken here, $(q^T C\gamma_5 q)q$, and the 
other is of the form $(q^T C q)\gamma_5 q$.
There are two reasons why we take the former.
It is well-known that 
the former operators strongly couple to the the ground-state 
baryons and reproduce experimental values well.
Therefore, it is suitable for investigation of the parity partner
of the ground state baryons.
Furthermore, the recent lattice calculation shows that these two operators give
the similar results for the negative-parity baryon spectrum
while the latter is more noisy \cite{SBO01}.

For baryons, two of three quark masses are taken to be
the same value as specified by the hopping parameter $\kappa_1$,
and the other quark mass is specified by $\kappa_2$.
This corresponds to taking the same value for $u, d$ current quark
masses as $m_u=m_d \equiv m_n$.
Then, the baryon masses are expressed as the function of two masses
$m_1$ and $m_2$, or equivalently of $\kappa_1$ and $\kappa_2$,
like $m_B(\kappa_1, \kappa_2)$.
In the source operator, each quark field is smeared with the
Gaussian function of width $\simeq$0.4 fm.

At large $t$ (and large $N_t - t$), the baryon correlators are
represented as
\begin{eqnarray}
G_B(t) &\equiv&  \sum_{\vec{x}} \langle
                      B(\vec{x},t) \bar{B}(\vec{x},0) \rangle
 \nonumber \\
  & & \hspace{-1.5cm}= \
    (1+\gamma_4) \left[ c_{B^+} \cdot e^{-tm_{B^+}}
           + b c_{B^-} \cdot e^{-(N_t - t) m_{B^-}} \right]
   \hspace{0.7cm}   \nonumber  \\
  & & \hspace{-1.3cm}
  + (1-\gamma_4) \left[ b c_{B^+} \cdot  e^{-(N_t - t)m_{B^+}}
             +  c_{B^-} \cdot e^{- t m_{B^-}} \right],
\label{eq:baryon_corr}
\end{eqnarray}
where $b=+1$ and $-1$ for the periodic and antiperiodic temporal
boundary conditions for the quark fields.
Since we adopt the standard Dirac representation for $\gamma$
matrices, the upper and
lower two components correspond to the first and second contributions
of Eq.~(\ref{eq:baryon_corr}).

Combining the parity-projected correlators
under two boundary conditions, one can single out the positive- 
and negative-parity baryon states with corresponding masses
$m_{B^+}$ and $m_{B^-}$, respectively, without contributions
from the backward propagating parity partners.
In practical simulation, however, we take a sufficient temporal extent
so that we can observe an enough range of plateau in effective mass
plot for extraction of mass in each parity channel,
and hence there is no advantage in computing correlators under two boundary
conditions except for the reduction of statistical fluctuation.
We obtain the baryon correlators at $\beta=5.75$ under two
boundary conditions, and compare the statistical fluctuations in
the parity-projected correlator and in not projected one.
We conclude that it is not worth doubling the computational cost
and hence adopt only the periodic boundary condition hereafter.
Instead, at each $\beta$
we obtain the correlators with the source at $t=N_t/2$ in addition
to ones with the source at $t=0$, and average them.
This is efficient to reduce the statistical errors for limited number
of configurations.

\begin{table*}
\caption{Typical interpolating operators for various hadrons.
For baryons, the contraction with the color index is omitted.
$C$ denotes the charge conjugate matrix.}
\label{tab:operator}
\begin{ruledtabular}
\begin{tabular}{ccc}
Meson 
& Pseudoscalar & $M(K)   = \bar{s} \gamma_5 u$   \\
& Vector       & $M_k(K^*) = \bar{s} \gamma_k u$ \\
\hline
Baryon
& Octet &
  $B_{\alpha}(\Sigma^0) = (C\gamma_5)_{\beta\gamma}[
      u_{\alpha}(d_{\beta}s_{\gamma} - s_{\beta}d_{\gamma}) 
    - d_{\alpha}(s_{\beta}u_{\gamma} - u_{\beta}s_{\gamma}) ]$ \\
& Octet ($\Lambda$) &
  $B_{\alpha}(\Lambda) = (C\gamma_5)_{\beta\gamma}[
      u_{\alpha}(d_{\beta}s_{\gamma} - s_{\beta}d_{\gamma})
    + d_{\alpha}(s_{\beta}u_{\gamma} - u_{\beta}s_{\gamma})
   -2 s_{\alpha}(u_{\beta}d_{\gamma} - d_{\beta}u_{\gamma}) ]$ \\
& Singlet  &
  $B_{\alpha}(\Lambda_1) = (C\gamma_5)_{\beta\gamma}[
      u_{\alpha}(d_{\beta}s_{\gamma} - s_{\beta}d_{\gamma}) 
    + d_{\alpha}(s_{\beta}u_{\gamma} - u_{\beta}s_{\gamma})
    + s_{\alpha}(u_{\beta}d_{\gamma} - d_{\beta}u_{\gamma}) ]$ \\
& Decuplet &
  $B_{\alpha k}(\Sigma^{*0}) = (C\gamma_k)_{\beta\gamma}[
      u_{\alpha}(d_{\beta}s_{\gamma} + s_{\beta}d_{\gamma}) 
    + d_{\alpha}(s_{\beta}u_{\gamma} + u_{\beta}s_{\gamma})
    + s_{\alpha}(u_{\beta}d_{\gamma} + d_{\beta}u_{\gamma}) ]$ \\
\end{tabular}
\end{ruledtabular}
\end{table*}

\subsection{lattice QCD results for hadron masses}

\begin{figure*}
\includegraphics[width=8.8cm]{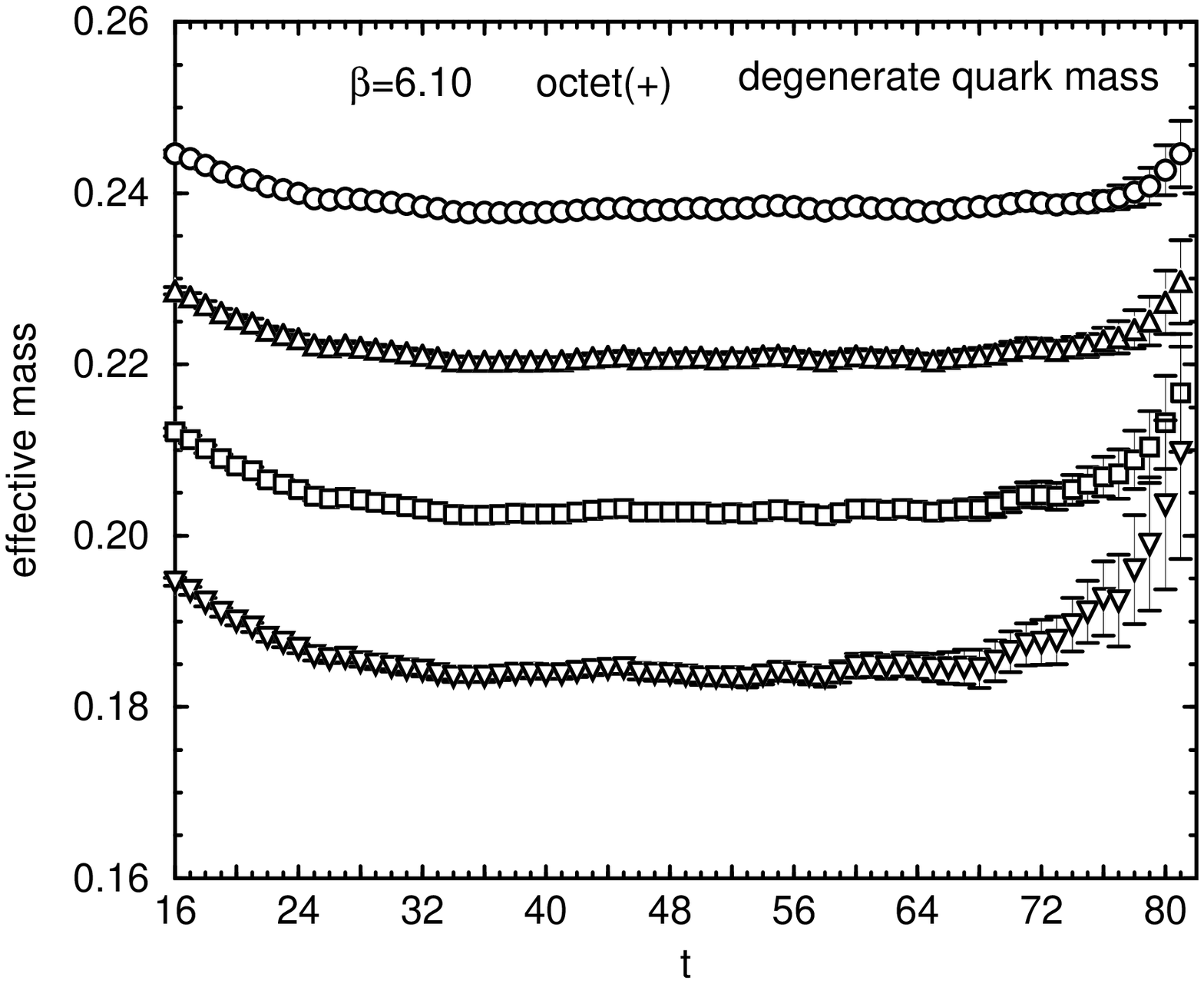}
\includegraphics[width=8.8cm]{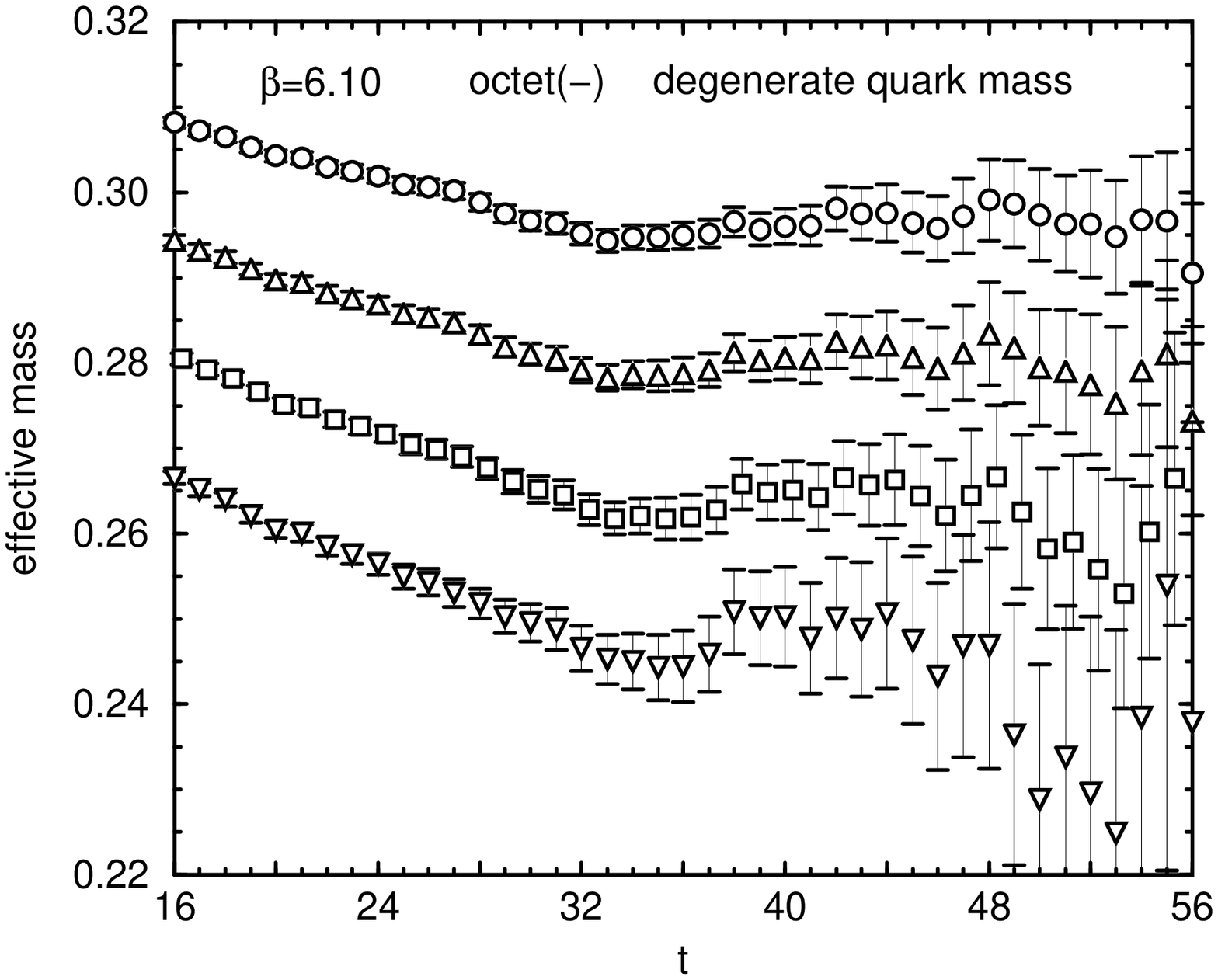}
\includegraphics[width=8.8cm]{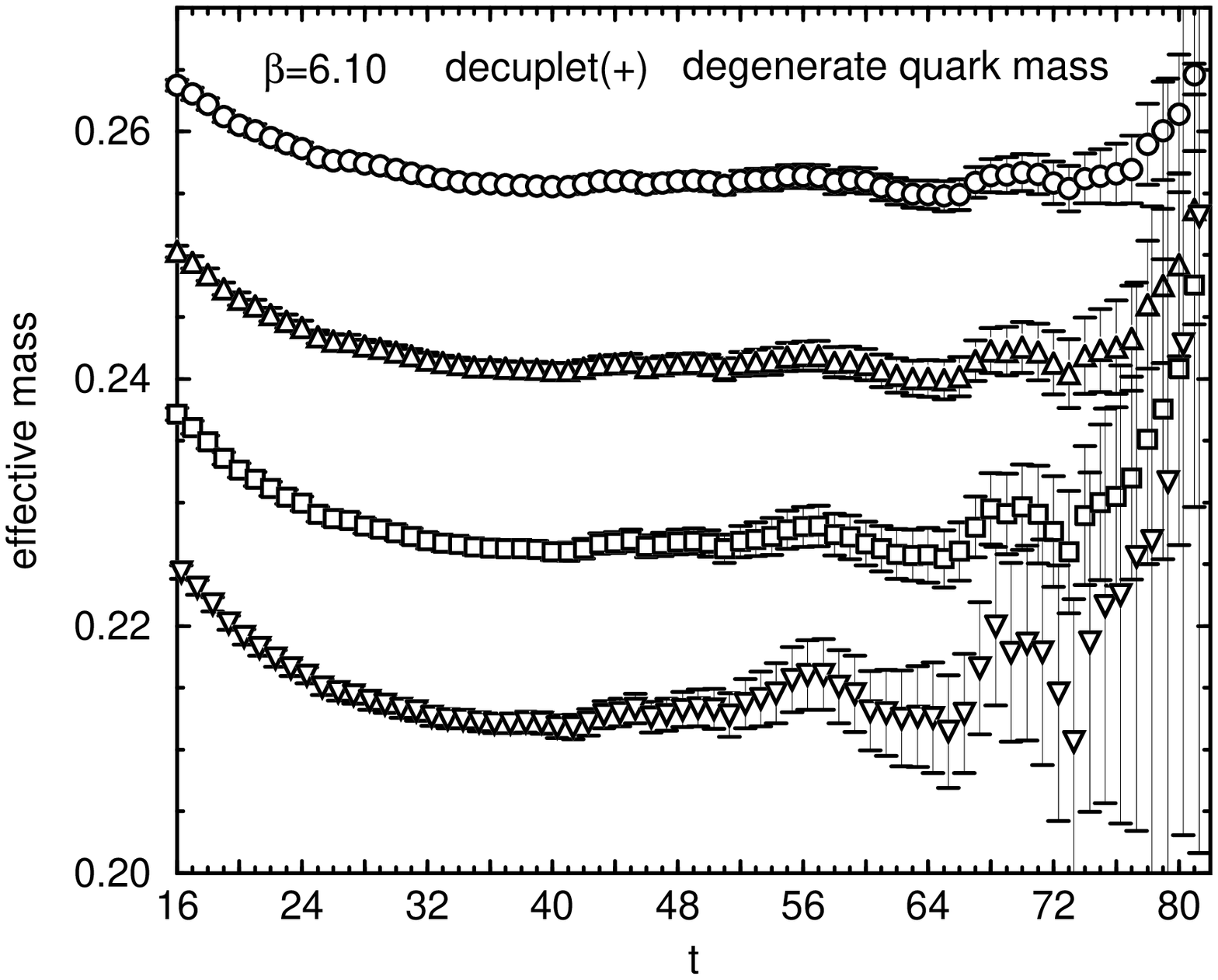}
\includegraphics[width=8.8cm]{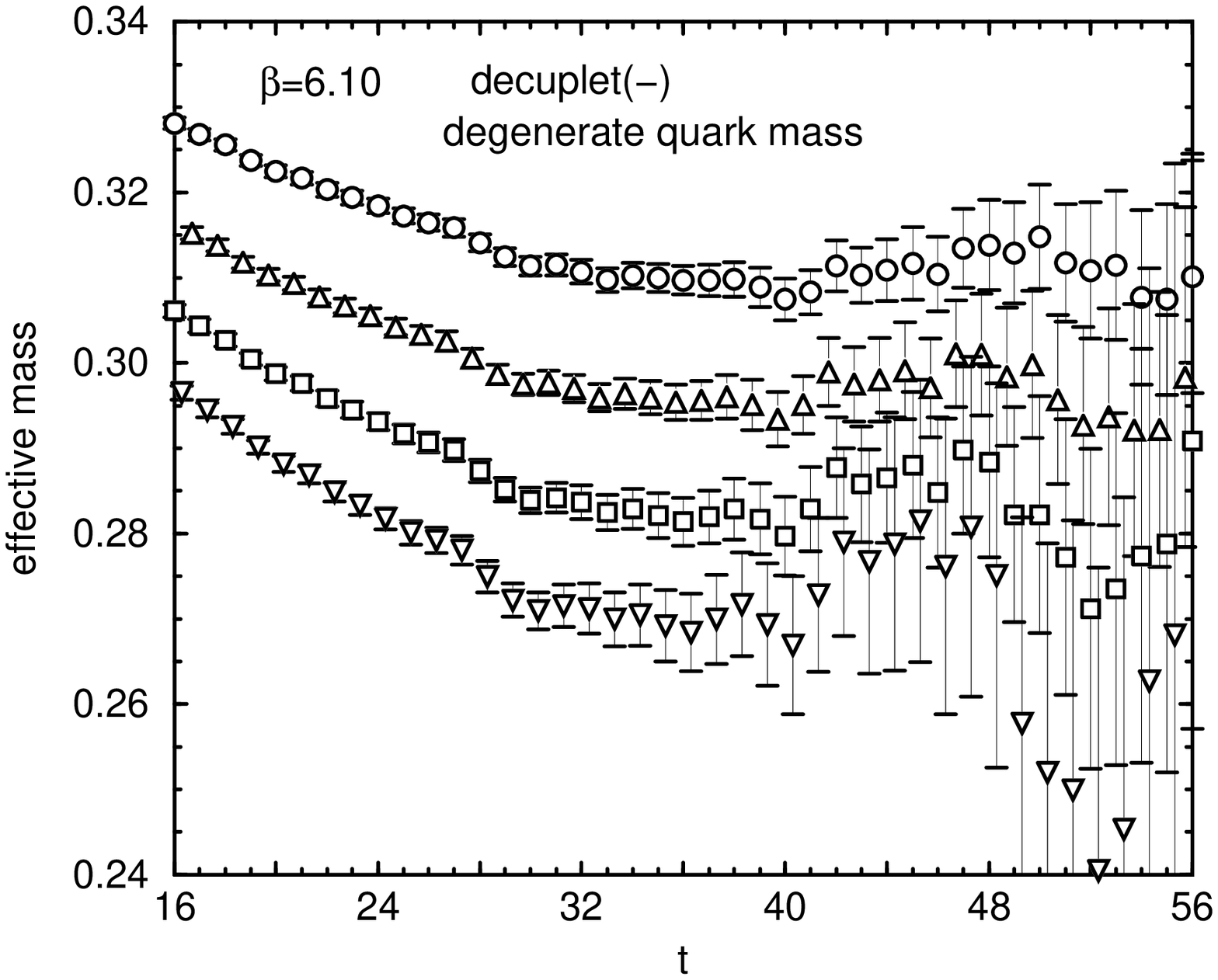}
\includegraphics[width=8.8cm]{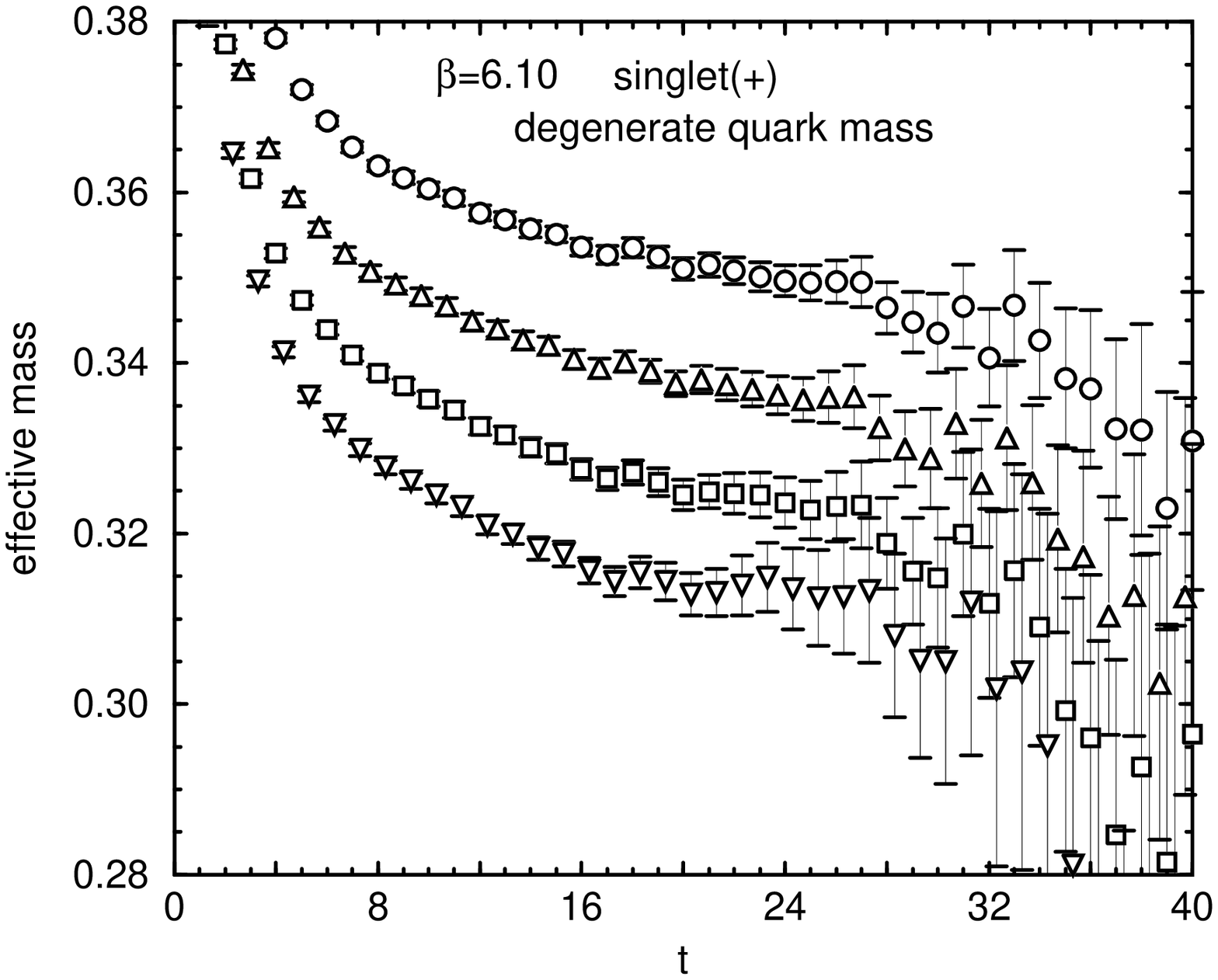}
\includegraphics[width=8.8cm]{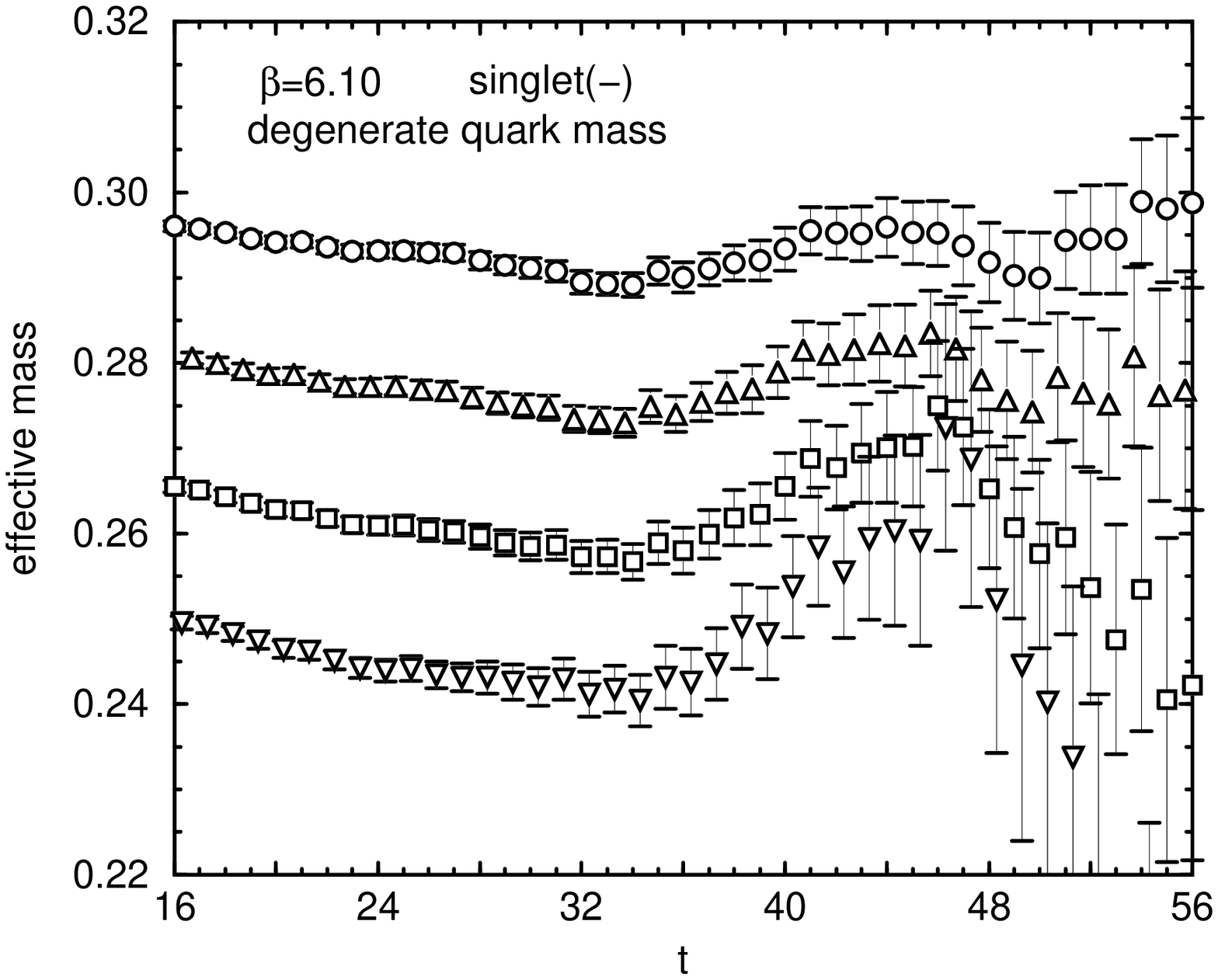}
\caption{
Effective mass plots for octet, decuplet and singlet baryon correlators
with degenerate quark masses at $\beta=6.10$.
The symbols correspond to
$\kappa=$0.1230, 0.1235, 0.1240 and 0.1245 from top to bottom in each 
figure.
The left figures are for the positive-parity baryons and the right for the
negative-parity ones.}
\label{fig:ep2G}
\end{figure*}

Fig.~\ref{fig:ep2G} shows
the effective mass plots for the baryon correlators at $\beta=6.10$.
The effective mass is defined without considering the contribution of the
associated parity partner propagating backward
from the source at $t=N_t$,
\begin{equation}
m_{\rm eff} = \ln\left( \frac{G_B(t)}{G_B(t+1)} \right).
\end{equation}
We observe that in the region where the effective mass exhibits 
a plateau,
the contribution of parity partner is sufficiently small.
In particular for the negative-parity baryon channels,
fine temporal lattice spacing seems to be helpful to specify
the region in which the ground state dominates.

\begin{figure}
\includegraphics[width=9cm]{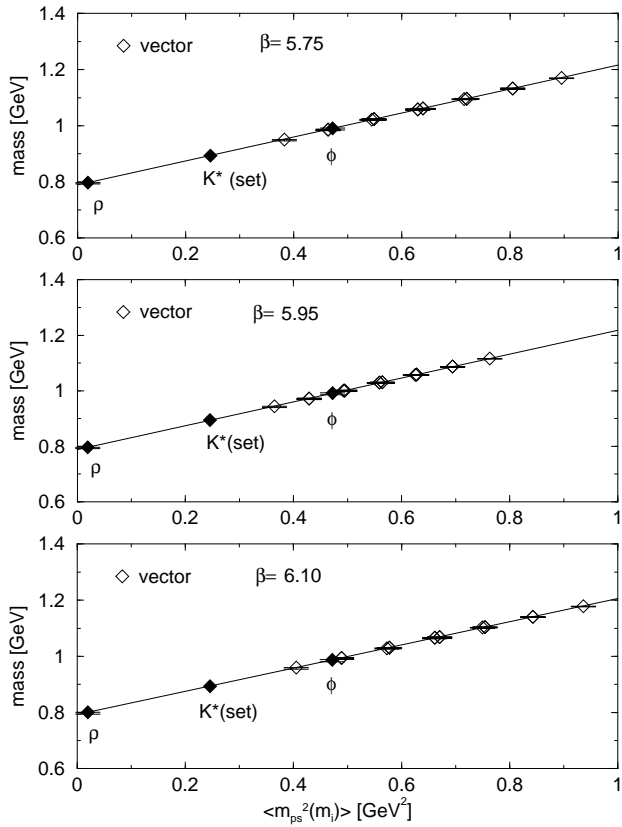}
\caption{The spectrum of the vector meson plotted against the
pseudoscalar meson mass squared $m_{\rm PS}^2$ in the physical unit. 
The open symbols denote the direct lattice data, 
and the filled symbols denote the results for the physical quark masses 
obtained from the lattice data with the linear chiral extrapolation.
}
\label{fig:meson}
\end{figure}

The meson correlator is fitted to the single hyperbolic cosine
form and analyzed independently of Ref. \cite{Aniso01b}.
The results are shown in Fig. \ref{fig:meson}.
and consistent with Ref. \cite{Aniso01b}.
For the baryons, we fit the data to a single exponential form.
The result is listed in Tables~\ref{tab:spectrum_E},
\ref{tab:spectrum_F} and \ref{tab:spectrum_G}.

\begin{table*}
\caption{
Baryon spectrum at $\beta=5.75$ in the temporal lattice unit.
When quark masses are degenerate as $\kappa_1=\kappa_2$,
the $\Sigma$-type and the $\Lambda$-type octet baryon correlators
become identical.}
\begin{ruledtabular}
\begin{tabular}{cccccccccc}
 & & \multicolumn{4}{c}{Positive-parity baryons}
   & \multicolumn{4}{c}{Negative-parity baryons} \\
$\kappa_1$ & $\kappa_2$ & 
$m_{\rm oct(\Sigma)}$ & $m_{\rm oct(\Lambda)}$ & $m_{\rm sing}$ &
$m_{\rm dec}$ &
$m_{\rm oct(\Sigma)}$ & $m_{\rm oct(\Lambda)}$ & $m_{\rm sing}$ &
$m_{\rm dec}$ \\
\hline
 0.1210& 0.1210& 0.4281( 9)& 0.4281( 9)& 0.6426(56)& 0.4606(16)&
                 0.5443(52)& 0.5443(52)& 0.5355(30)& 0.5630(54)\\
 0.1210& 0.1220& 0.4171(10)& 0.4186(10)& 0.6349(59)& 0.4518(17)&
                 0.5355(56)& 0.5356(57)& 0.5265(32)& 0.5548(59)\\
 0.1210& 0.1230& 0.4059(10)& 0.4091(10)& 0.6267(64)& 0.4433(19)&
                 0.5271(61)& 0.5270(64)& 0.5175(34)& 0.5467(66)\\
 0.1210& 0.1240& 0.3946(11)& 0.3998(11)& 0.6184(69)& 0.4352(21)&
                 0.5189(68)& 0.5178(78)& 0.5089(37)& 0.5386(78)\\
\hline
 0.1220& 0.1210& 0.4083(10)& 0.4066(10)& 0.6275(63)& 0.4430(19)&
                 0.5269(61)& 0.5269(60)& 0.5173(33)& 0.5465(65)\\
 0.1220& 0.1220& 0.3970(11)& 0.3970(11)& 0.6206(68)& 0.4341(20)&
                 0.5181(66)& 0.5181(66)& 0.5080(35)& 0.5382(71)\\
 0.1220& 0.1230& 0.3855(11)& 0.3874(11)& 0.6132(74)& 0.4256(22)&
                 0.5095(73)& 0.5091(75)& 0.4989(38)& 0.5300(79)\\
 0.1220& 0.1240& 0.3737(12)& 0.3780(12)& 0.6055(81)& 0.4176(25)&
                 0.5012(82)& 0.4994(90)& 0.4899(41)& 0.5217(93)\\
\hline
 0.1230& 0.1210& 0.3885(11)& 0.3845(11)& 0.6116(74)& 0.4259(22)&
                 0.5091(77)& 0.5099(74)& 0.4990(38)& 0.5302(81)\\
 0.1230& 0.1220& 0.3769(12)& 0.3747(12)& 0.6063(81)& 0.4171(24)&
                 0.5001(83)& 0.5008(81)& 0.4895(41)& 0.5218(89)\\
 0.1230& 0.1230& 0.3650(12)& 0.3650(12)& 0.6004(89)& 0.4085(27)&
                 0.4913(92)& 0.4913(92)& 0.4801(44)& 0.513(10) \\
 0.1230& 0.1240& 0.3527(13)& 0.3554(13)& 0.5939(99)& 0.4004(32)&
                 0.483(11) & 0.481(11) & 0.4707(48)& 0.505(12) \\
\hline
 0.1240& 0.1210& 0.3689(13)& 0.3612(13)& 0.5936(89)& 0.4099(30)&
                 0.489(11) & 0.4928(98)& 0.4805(46)& 0.514(11) \\
 0.1240& 0.1220& 0.3569(14)& 0.3512(14)& 0.591(10) & 0.4010(33)&
                 0.480(12) & 0.483(11) & 0.4708(49)& 0.505(12) \\
 0.1240& 0.1230& 0.3445(15)& 0.3414(15)& 0.588(11) & 0.3922(37)&
                 0.470(13) & 0.473(12) & 0.4610(53)& 0.496(14) \\
 0.1240& 0.1240& 0.3317(16)& 0.3317(16)& 0.585(13) & 0.3838(45)&
                 0.460(15) & 0.460(15) & 0.4509(58)& 0.486(16) \\
\hline
 \multicolumn{2}{c}{fit range}
     &   24--40 &   24--40 &   12--20 &   28--40 
     &   20--32 &   20--32 &   16--24 &   20--32
\end{tabular}
\end{ruledtabular}
\label{tab:spectrum_E}
\end{table*}

\begin{table*}
\caption{
The same results as Table~\ref{tab:spectrum_E} for $\beta=5.95$.}
\begin{ruledtabular}
\begin{tabular}{cccccccccc}
 & & \multicolumn{4}{c}{Positive-parity baryons}
   & \multicolumn{4}{c}{Negative-parity baryons} \\
$\kappa_1$ & $\kappa_2$ & 
$m_{\rm oct(\Sigma)}$ & $m_{\rm oct(\Lambda)}$ & $m_{\rm sing}$ &
$m_{\rm dec}$ &
$m_{\rm oct(\Sigma)}$ & $m_{\rm oct(\Lambda)}$ & $m_{\rm sing}$ &
$m_{\rm dec}$ \\
\hline
 0.1230& 0.1230& 0.2785( 7)& 0.2785( 7)& 0.4279(27)& 0.3041(10)&
                 0.3537(28)& 0.3537(28)& 0.3460(32)& 0.3771(36)\\
 0.1230& 0.1235& 0.2722( 7)& 0.2732( 7)& 0.4240(28)& 0.2993(10)&
                 0.3482(30)& 0.3487(30)& 0.3407(33)& 0.3728(39)\\
 0.1230& 0.1240& 0.2658( 7)& 0.2680( 8)& 0.4202(30)& 0.2945(11)&
                 0.3428(32)& 0.3441(34)& 0.3355(36)& 0.3690(43)\\
 0.1230& 0.1245& 0.2592( 8)& 0.2628( 8)& 0.4165(34)& 0.2899(12)&
                 0.3374(35)& 0.3406(40)& 0.3306(40)& 0.3661(48)\\
\hline
 0.1235& 0.1230& 0.2675( 7)& 0.2664( 7)& 0.4203(30)& 0.2944(11)&
                 0.3434(33)& 0.3430(32)& 0.3354(35)& 0.3686(42)\\
 0.1235& 0.1235& 0.2610( 8)& 0.2610( 8)& 0.4168(33)& 0.2896(12)&
                 0.3378(35)& 0.3378(35)& 0.3299(38)& 0.3643(45)\\
 0.1235& 0.1240& 0.2545( 8)& 0.2557( 8)& 0.4133(36)& 0.2848(12)&
                 0.3322(38)& 0.3330(39)& 0.3246(41)& 0.3604(49)\\
 0.1235& 0.1245& 0.2477( 8)& 0.2504( 9)& 0.4099(40)& 0.2801(14)&
                 0.3268(42)& 0.3291(46)& 0.3195(45)& 0.3575(56)\\
\hline
 0.1240& 0.1230& 0.2564( 8)& 0.2538( 8)& 0.4129(36)& 0.2849(13)&
                 0.3332(40)& 0.3324(38)& 0.3247(41)& 0.3608(50)\\
 0.1240& 0.1235& 0.2498( 8)& 0.2484( 8)& 0.4099(40)& 0.2800(13)&
                 0.3275(42)& 0.3270(41)& 0.3192(44)& 0.3565(54)\\
 0.1240& 0.1240& 0.2430( 9)& 0.2430( 9)& 0.4070(44)& 0.2751(14)&
                 0.3218(46)& 0.3218(46)& 0.3136(48)& 0.3527(60)\\
 0.1240& 0.1245& 0.2359( 9)& 0.2375(10)& 0.4041(50)& 0.2704(16)&
                 0.3162(52)& 0.3175(55)& 0.3083(54)& 0.3498(68)\\
\hline
 0.1245& 0.1230& 0.2451( 9)& 0.2404( 9)& 0.4054(46)& 0.2754(15)&
                 0.3239(53)& 0.3225(49)& 0.3143(52)& 0.3548(64)\\
 0.1245& 0.1235& 0.2383(10)& 0.2349( 9)& 0.4035(51)& 0.2705(16)&
                 0.3180(57)& 0.3168(53)& 0.3085(56)& 0.3507(69)\\
 0.1245& 0.1240& 0.2312(10)& 0.2294(10)& 0.4016(58)& 0.2656(18)&
                 0.3122(63)& 0.3112(60)& 0.3028(61)& 0.3470(77)\\
 0.1245& 0.1245& 0.2237(11)& 0.2237(11)& 0.4000(68)& 0.2609(19)&
                 0.3065(71)& 0.3065(71) & 0.2972(70)& 0.3444(89)\\
\hline
 \multicolumn{2}{c}{fit range}
     &   28--56 &   28--56 &   16--24 &   28--58 
     &   26--44 &   26--44 &   26--44 &   26--44  \\
\end{tabular}
\end{ruledtabular}
\label{tab:spectrum_F}
\end{table*}

\begin{table*}
\caption{
The same results as Table~\ref{tab:spectrum_E} for $\beta=6.10$.}
\begin{ruledtabular}
\begin{tabular}{cccccccccc}
 & & \multicolumn{4}{c}{Positive-parity baryons}
   & \multicolumn{4}{c}{Negative-parity baryons} \\
$\kappa_1$ & $\kappa_2$ & 
$m_{\rm oct(\Sigma)}$ & $m_{\rm oct(\Lambda)}$ & $m_{\rm sing}$ &
$m_{\rm dec}$ &
$m_{\rm oct(\Sigma)}$ & $m_{\rm oct(\Lambda)}$ & $m_{\rm sing}$ &
$m_{\rm dec}$ \\
\hline
 0.1230& 0.1230& 0.2382( 4)& 0.2382( 4)& 0.3466(30)& 0.2560( 7)&
                 0.2967(23)& 0.2967(23)& 0.2925(20)& 0.3103(26)\\
 0.1230& 0.1235& 0.2320( 5)& 0.2329( 5)& 0.3417(32)& 0.2511(08)&
                 0.2911(25)& 0.2919(25)& 0.2875(21)& 0.3059(28)\\
 0.1230& 0.1240& 0.2257( 5)& 0.2276( 5)& 0.3365(35)& 0.2463(08)&
                 0.2854(27)& 0.2874(27)& 0.2827(22)& 0.3017(32)\\
 0.1230& 0.1245& 0.2193( 5)& 0.2223( 5)& 0.3311(39)& 0.2417(09)&
                 0.2796(29)& 0.2835(33)& 0.2783(25)& 0.2981(38)\\
\hline
 0.1235& 0.1230& 0.2271( 5)& 0.2262( 5)& 0.3370(35)& 0.2462( 8)&
                 0.2867(26)& 0.2858(26)& 0.2826(22)& 0.3014(31)\\
 0.1235& 0.1235& 0.2208( 5)& 0.2208( 5)& 0.3325(38)& 0.2413(09)&
                 0.2809(28)& 0.2809(28)& 0.2776(24)& 0.2969(34)\\
 0.1235& 0.1240& 0.2143( 5)& 0.2154( 5)& 0.3277(42)& 0.2365(10)&
                 0.2750(31)& 0.2762(31)& 0.2728(26)& 0.2926(38)\\
 0.1235& 0.1245& 0.2076( 6)& 0.2101( 6)& 0.3226(48)& 0.2319(11)&
                 0.2689(34)& 0.2722(38)& 0.2683(29)& 0.2889(46)\\
\hline
 0.1240& 0.1230& 0.2160( 5)& 0.2137( 5)& 0.3268(42)& 0.2366(10)&
                 0.2768(32)& 0.2746(31)& 0.2729(26)& 0.2929(39)\\
 0.1240& 0.1235& 0.2095( 6)& 0.2082( 5)& 0.3232(47)& 0.2317(11)&
                 0.2709(34)& 0.2695(34)& 0.2680(28)& 0.2883(43)\\
 0.1240& 0.1240& 0.2028( 6)& 0.2028( 6)& 0.3193(53)& 0.2270(12)&
                 0.2646(37)& 0.2646(37)& 0.2631(31)& 0.2840(49)\\
 0.1240& 0.1245& 0.1958( 6)& 0.1974( 6)& 0.3150(62)& 0.2225(13)&
                 0.2581(42)& 0.2604(45)& 0.2586(35)& 0.2803(59)\\
\hline
 0.1245& 0.1230& 0.2049( 6)& 0.2006( 6)& 0.3151(55)& 0.2276(12)&
                 0.2680(44)& 0.2628(42)& 0.2638(33)& 0.2855(56)\\
 0.1245& 0.1235& 0.1982( 7)& 0.1950( 6)& 0.3133(63)& 0.2228(14)&
                 0.2617(48)& 0.2575(45)& 0.2588(36)& 0.2809(63)\\
 0.1245& 0.1240& 0.1912( 7)& 0.1894( 7)& 0.3114(74)& 0.2182(15)&
                 0.2551(53)& 0.2523(50)& 0.2540(40)& 0.2765(72)\\
 0.1245& 0.1245& 0.1839( 8)& 0.1839( 8)& 0.3094(91)& 0.2139(18)&
                 0.2479(61)& 0.2479(61)& 0.2493(46)& 0..2727(88)\\
\hline
 \multicolumn{2}{c}{fit range}
     &   40--64 &   40--64 &   24--36 &   44--64 
     &   36--52 &   36--52 &   32--52 &   36--52  \\
\end{tabular}
\end{ruledtabular}
\label{tab:spectrum_G}
\end{table*}

Following Ref. \cite{Aniso01b}, we extrapolate the hadron masses
to the chiral limit in terms of the pseudoscalar
meson mass squared, instead of $1/\kappa$.
The assumed relation between PS meson mass and quark mass is
\begin{equation}
 m_{PS}^2(m_1, m_2) = B \cdot (m_1 + m_2),
\end{equation}
then for the degenerate quark masses, $m_1=m_2$,
$m_{PS}^2 = 2 B m_1$ holds.
Instead of $m_i$ ($i$=1,2), one can extrapolate other hadron masses
in term of $m_{PS}(m_i,m_i)^2$ to the chiral limit.

In our calculation for baryons, two of quark masses are taken to be
the same value, $m_1$, and the other quark mass $m_2$ is
taken to be an independent value.
Then, the baryon masses are expressed as the function of $m_1$ and
$m_2$ like $m_B(m_1,m_2)$, and therefore they 
are to be depicted on the $(m_1,m_2)$ plane. 
However, the result for the baryon masses seem to
be well described with the linear relation,
\begin{equation}
 m_{B} (m_1, m_2, m_3) = m_B(0,0,0) + B_B \cdot (2 m_1 + m_2).
\end{equation}
Therefore, we fit the baryon mass data to the linear form in
the sum of corresponding PS meson masses squared.
The vector meson is also fitted to a linear function in 
$m_1 + m_2$.

Note here that, in quenched QCD, a non-analyticity appears in the chiral 
extrapolation near the chiral limit~\cite{Lei20}. 
For nucleons, it is reported that the departure from the simple
chiral extrapolation is observed for $m_\pi< 400$ MeV~\cite{DL03}.
Although we have to keep this effect in our mind, 
we do not argue it here because there is no distinct behavior 
for both the positive- and negative-parity baryons in our calculation
with $m_\pi>600$MeV.

\begin{figure*}
\includegraphics[width=8.8cm]{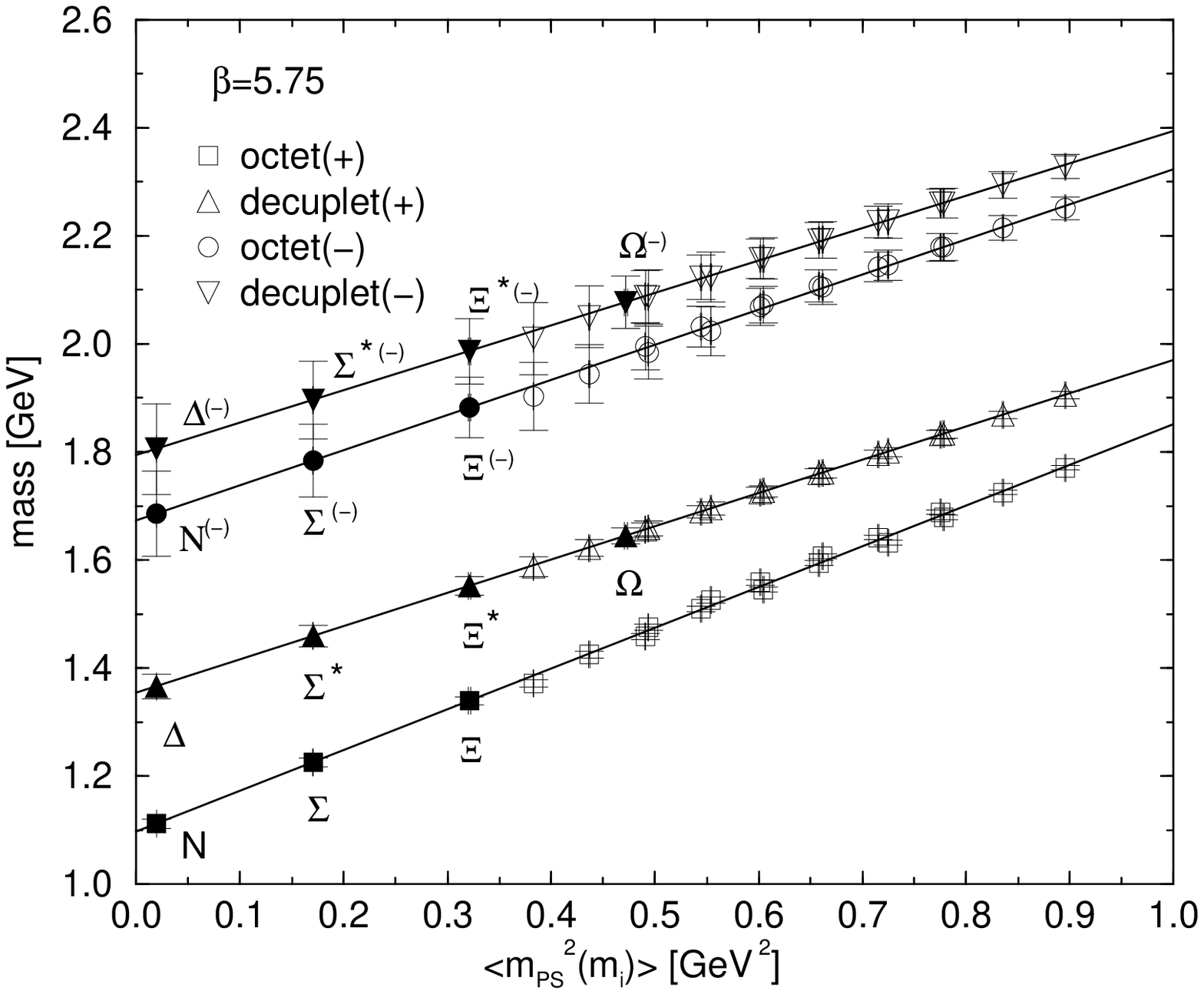}
\includegraphics[width=8.8cm]{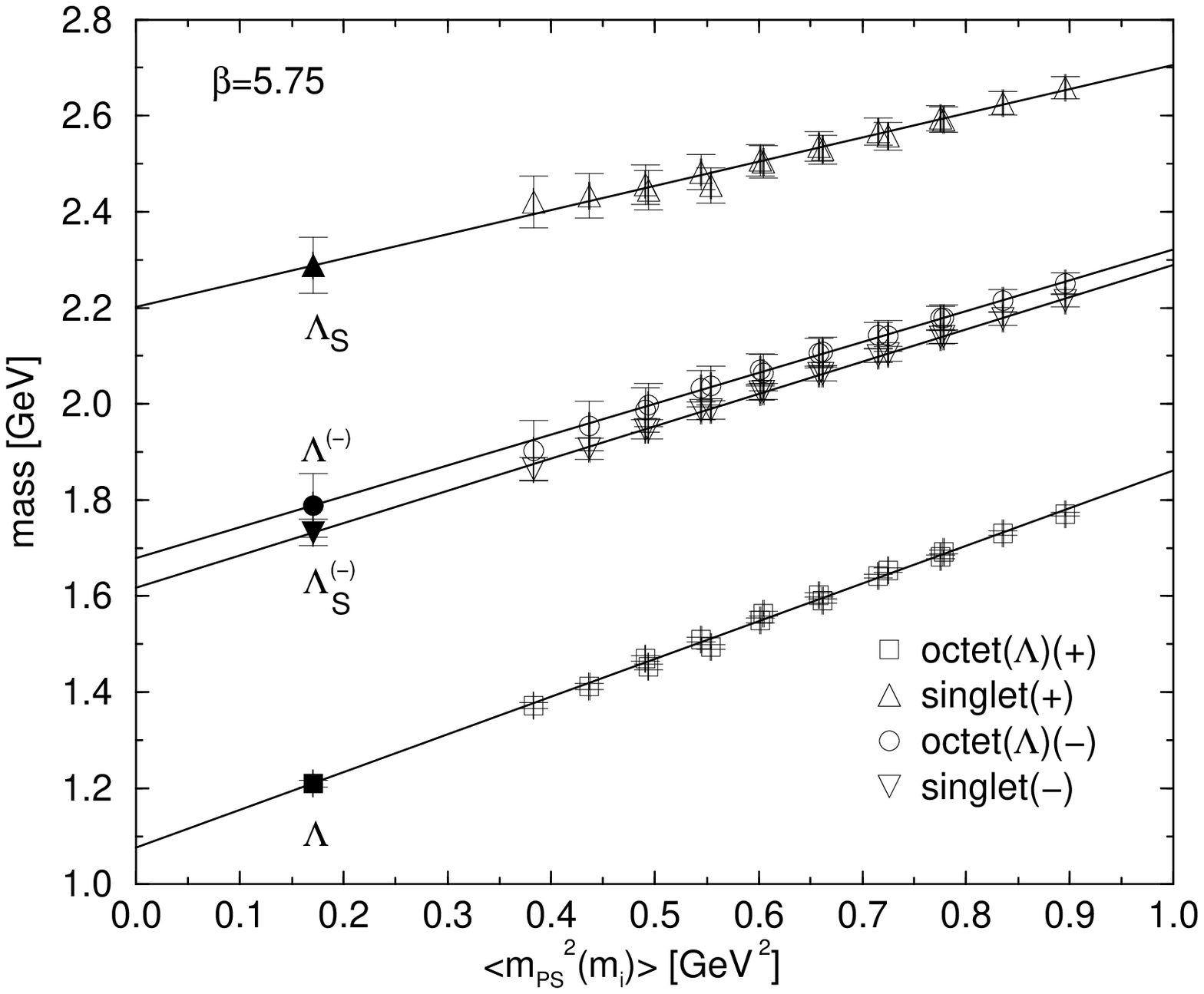}
\includegraphics[width=8.8cm]{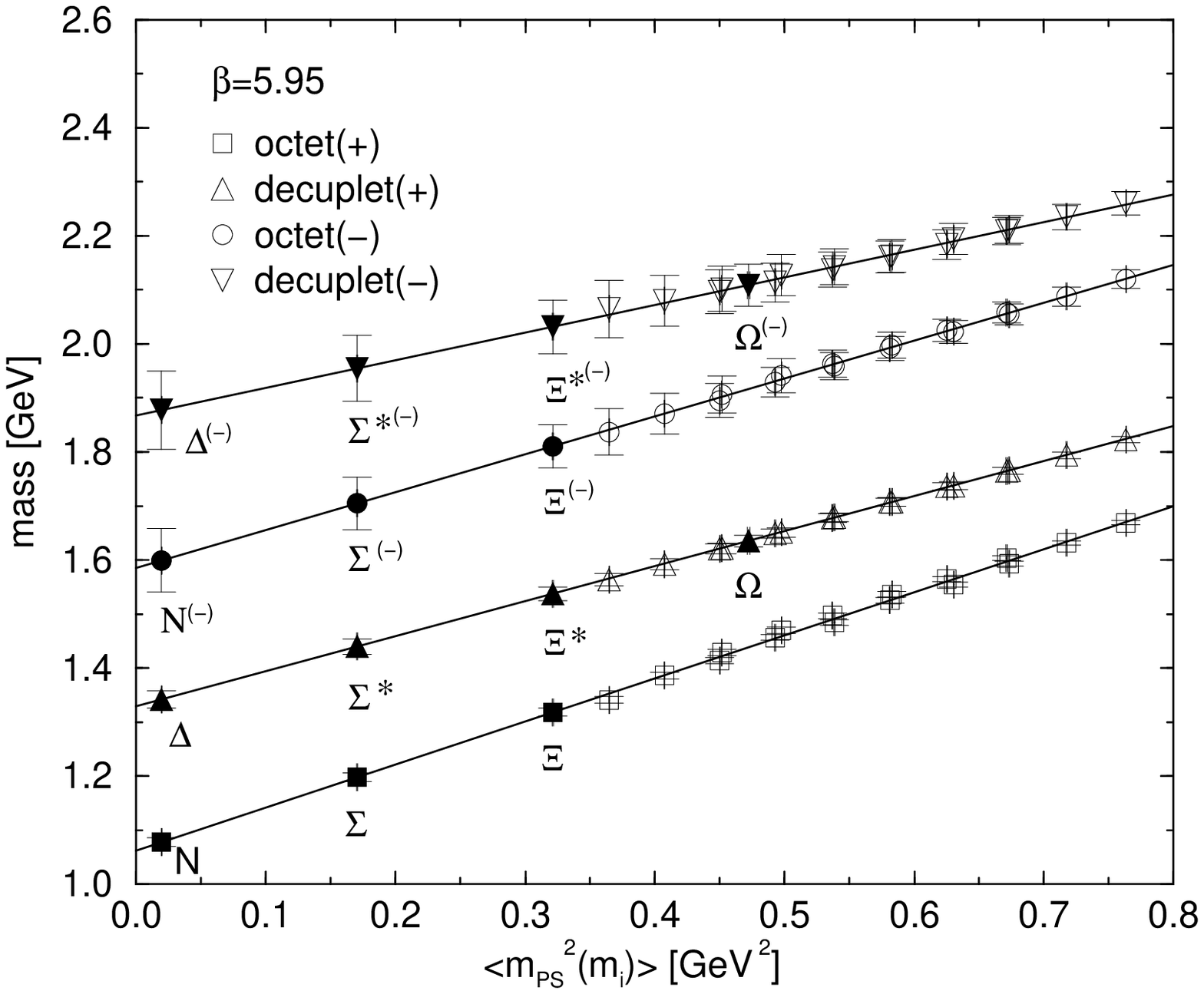}
\includegraphics[width=8.8cm]{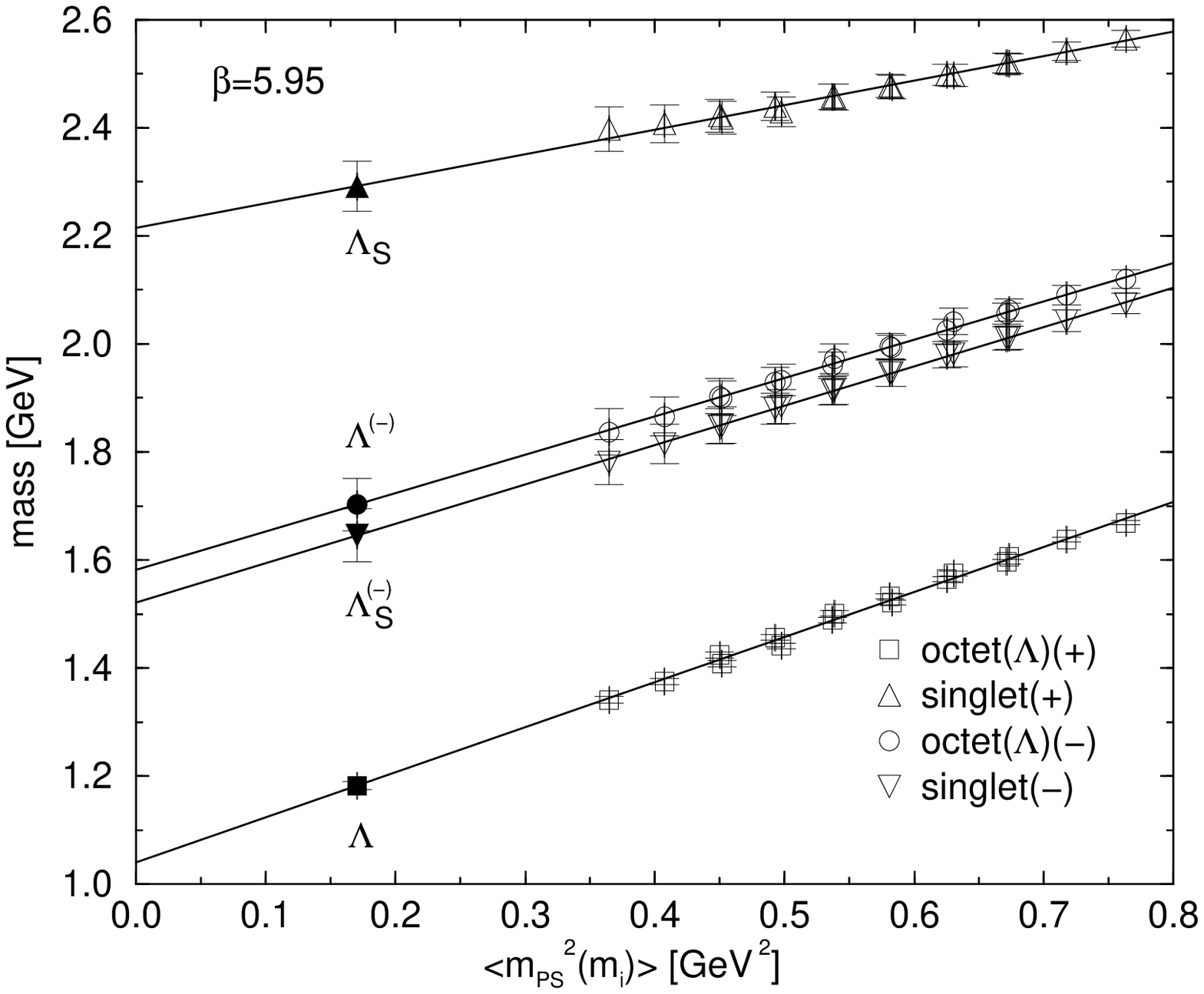}
\includegraphics[width=8.8cm]{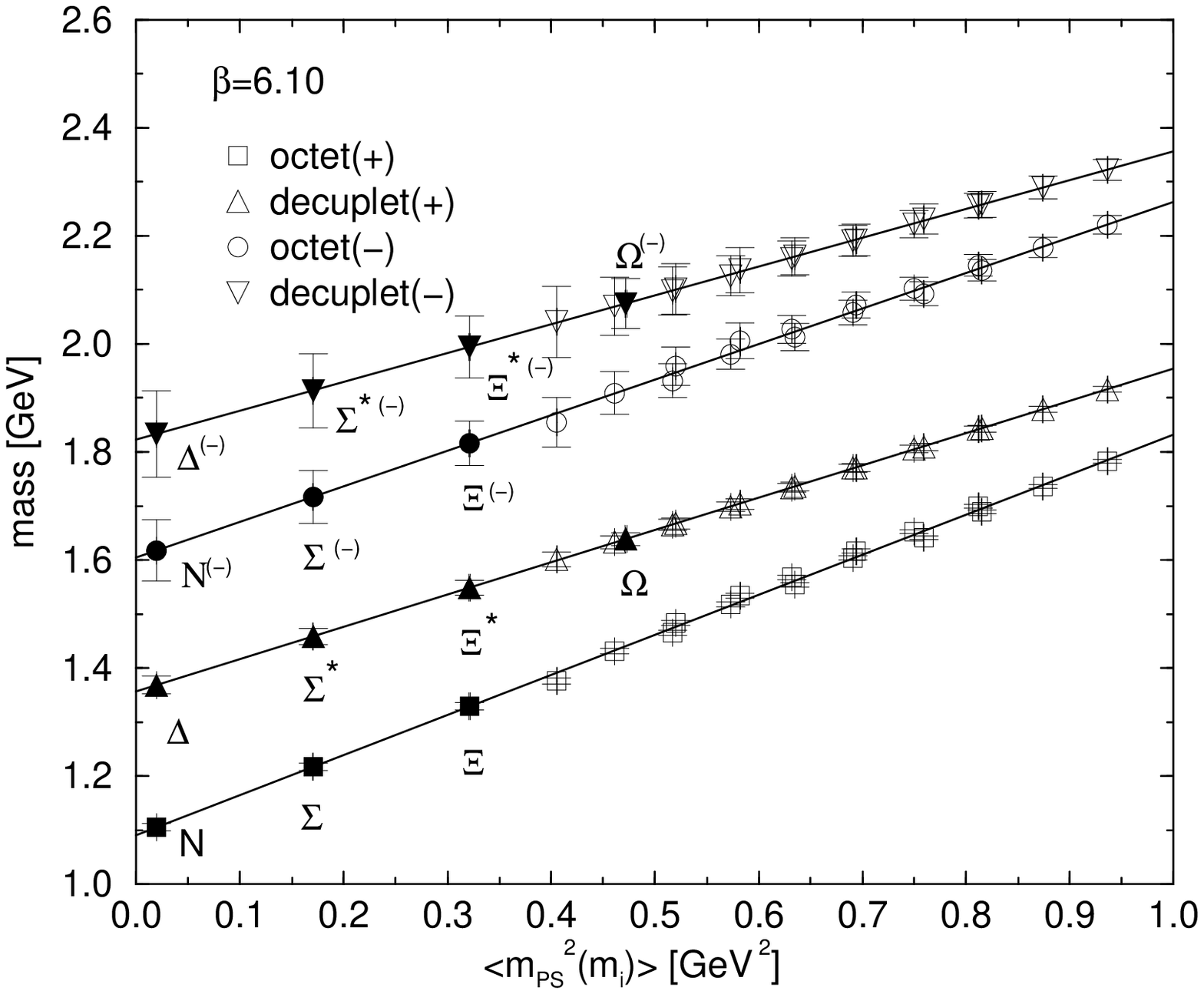}
\includegraphics[width=8.8cm]{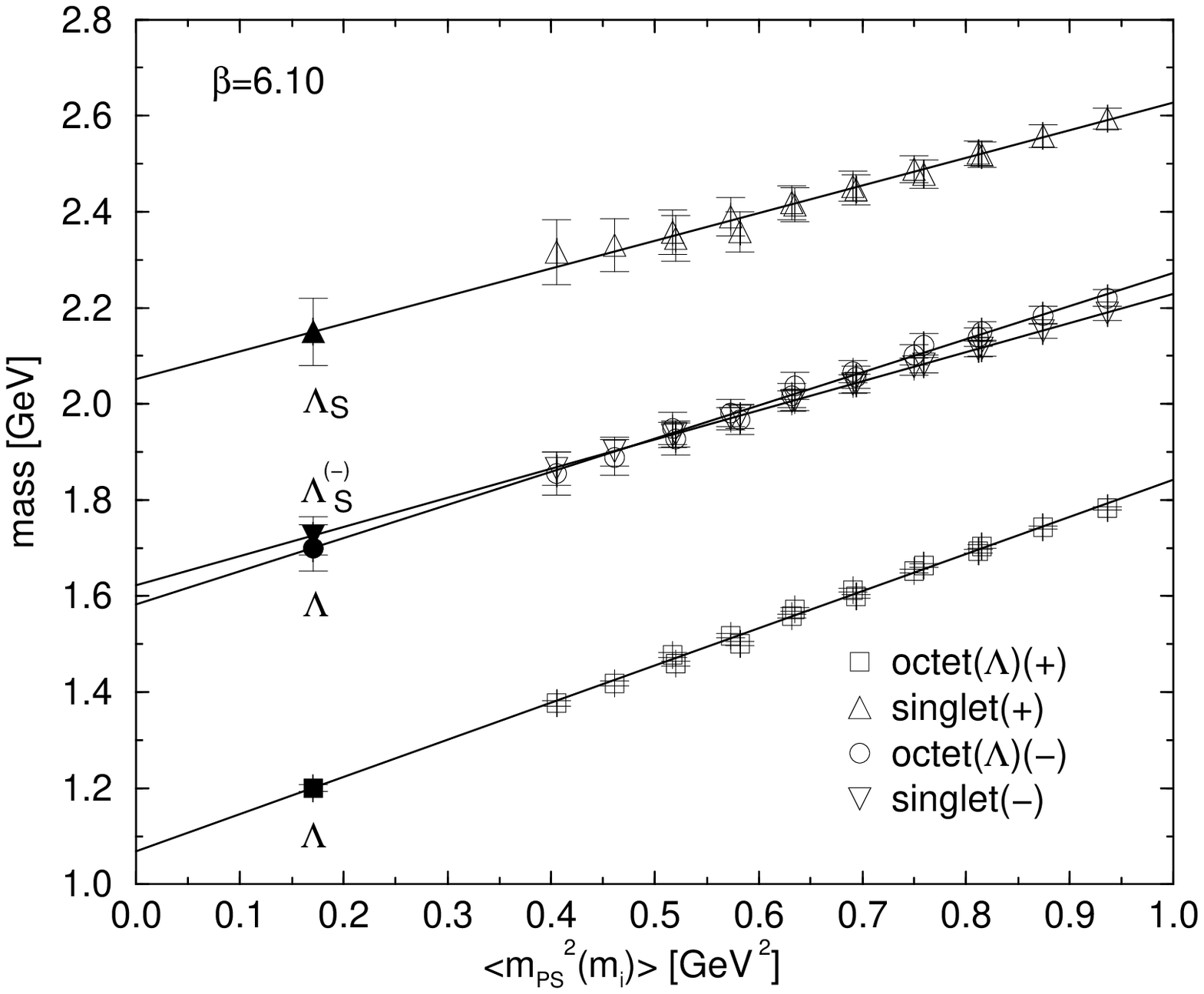}
\caption{
The spectra of positive- and negative-parity baryons plotted against
the pseudoscalar meson mass squared $m_{\rm PS}^2$.
For each $\beta$, the octet and the decuplet baryons are shown in the
left panel
and octet($\Lambda$) and singlet baryons in the right panel.
The horizontal axis denotes the averaged pseudoscalar mass square.
The open symbols denote the direct lattice data, 
and the filled symbols denote the results for the physical quark masses 
obtained from the lattice data with the linear chiral extrapolation.
}
\label{fig:spectrum1}
\end{figure*}

The results of fits for baryons are shown in 
Fig.~\ref{fig:spectrum1} for each lattice.
The horizontal axis is the averaged pseudoscalar meson mass squared,
\begin{equation}
 \langle m_{PS}^2(m_i) \rangle
 = \frac{1}{N_q} \sum_{i=1}^{N_q} m_{PS}^2(m_i,m_i)
 = \frac{1}{N_q} \sum_{i=1}^{N_q} 2B m_i
\end{equation}
with $N_q=3$ for baryons.
The results of fits are displayed as the solid lines in
these figures.
The linear relation seems to hold well.

As stated in Section~\ref{sec:simulation}, 
we determine the scale $a_{\tau}^{-1}$ 
through the $K^*$ meson mass.
The physical ($u$, $d$) and $s$ quark masses are
determined with the $\pi$ and $K$ meson masses.
The scales of the vertical and horizontal axes in Fig.~\ref{fig:spectrum1}
are set in this way.
The corresponding hadron masses for the physical quark masses are listed 
in Table~\ref{tab:spectrum_phys2}.
These results for the meson masses and positive-parity baryon masses
are consistent with those obtained in Ref. \cite{Aniso01b}.

\begin{table}
\caption{
The Hadron spectrum expressed in the unit of GeV at the physical quark masses.
The $K^*$ meson mass is used for the determination of the scale unit $a_\tau$.}
\begin{ruledtabular}
\begin{tabular}{lccc}
   & $\beta=5.75$ & $\beta=5.95$ & $\beta=6.10$ \\
\hline
$\rho$          & 0.7973(52)& 0.7965(53)& 0.8005(65) \\
$K^*$           & 0.8939    &  0.8939   & 0.8939     \\
$\phi$          & 0.9905(55)& 0.9913(53)& 0.9873(66) \\
\hline
$N$             & 1.1118(84)& 1.0781(81)& 1.1055(72)\\
$\Lambda$       & 1.2095(75)& 1.1825(75)& 1.2002(67)\\
$\Sigma$        & 1.2256(76)& 1.1982(76)& 1.2173(67)\\
$\Xi$           & 1.3393(71)& 1.3183(75)& 1.3291(67)\\
\hline
$\Delta$        & 1.366(23) & 1.342(16) & 1.3685(17) \\
$\Sigma^*$      & 1.459(20) & 1.440(14) & 1.4586(15) \\
$\Xi^*$         & 1.552(17) & 1.538(12) & 1.5486(13) \\
$\Omega$        & 1.645(15) & 1.635(11) & 1.6387(12) \\
\hline
$N^{(-)}$       & 1.686(79) & 1.599(59) & 1.618(57) \\
$\Sigma^{(-)}$  & 1.784(67) & 1.705(49) & 1.717(49) \\
$\Xi^{(-)}$     & 1.882(56) & 1.810(40) & 1.816(42) \\
\hline
$\Lambda^{(-)}_{\rm oct}$  
                & 1.788(66) & 1.703(48) & 1.700(49) \\
$\Lambda^{(-)}_{\rm sing}$
                & 1.732(28) & 1.646(49) & 1.725(39) \\
\hline
$\Delta^{(-)}$  & 1.806(84) & 1.877(73) & 1.833(80) \\
$\Sigma^{*(-)}$ & 1.896(72) & 1.955(61) & 1.913(69) \\
$\Xi^{*(-)}$    & 1.986(60) & 2.032(50) & 1.994(57) \\
$\Omega^{(-)}$  & 2.077(49) & 2.109(39) & 2.074(46) \\
\hline
$\Lambda^{(+)}_{\rm sing}$
                & 2.288(58) & 2.292(46) & 2.150(70) \\
\end{tabular}
\end{ruledtabular}
\label{tab:spectrum_phys2}
\end{table}

\subsection{Systematic errors}

Finally before discussing physical implications of our numerical results,
we briefly comment on the systematic uncertainties.

\begin{itemize}

\item {\it Anisotropy (calibration)}

According to the detailed inspection given in Ref. \cite{Aniso01b},
the 2\% uncertainty in $\gamma_F$ causes uncertainties in hadron masses
at 1\% level.
Although the effect of uncertainty coming from anisotropy on the negative
parity baryon masses is unknown,
its size is expected to be the same level
as for the positive-parity baryons and smaller than their
statistical errors.
Therefore, we do not perform a detailed analysis of 
the calibration uncertainty here.
The uncertainty in $\gamma_G$, the gluonic anisotropy parameter,
is also kept within 1\% level, and hence for the same reason as for
$\gamma_F$, we do not argue its effect on the negative-parity
baryon masses.

\item {\it Finite volume effects}

Since the excited baryons may have larger spatial extent than
the ground state baryons, they may seriously suffer from the finite
size effects.
Our present three lattices, however, have almost the same size
($\sim$ 2 fm), and we cannot examine the finite volume effects
on these lattices.
In Ref.~\cite{Goc01}, the finite volume effect on the negative-parity
baryon masses was evaluated as 5\% by comparing the masses on
lattices with volume sizes 1.5fm and 2.2fm (1.6fm and 2.1fm)
at $\beta=6.0$ (6.2) on quenched isotropic lattices.
This amount of finite size effect may also exist in our results,
while our lattice volumes are close to the larger ones in
Ref.~\cite{Goc01}.

\item {\it Lattice discretization error}

Table \ref{tab:spectrum_phys2} shows the baryon masses on each lattice.
We find it hard to take the continuum limit even for the
ground-state baryons and mesons, since only the three $\beta$'s are
taken here and their behavior is not so smooth that one can apply
a simple extrapolation.
In addition, such fluctuating behavior of data may be large due to
the lack of statistics, genuine discretization errors would not
be negligible.
In the range of lattice cutoff 1--2 GeV,
the fluctuation of masses is at most about 5\%, except for the
case of $\Lambda_{\rm sing}^{+}$ for which 7\% deviation is found.
This gives us a hint on the potential size of the discretization
errors.
We also note that
Ref.~\cite{Aniso01b} examined how the $O(\alpha a)$ and $O(a^2)$
discretization effects decrease as $\beta$ increases in the meson sectors,
and found that those in the calibration of $\gamma_F$
are sufficiently reduced already at $\beta=6.10$.
For these reason we discuss physical consequences of our result
mainly based on the data of $\beta=6.10$ lattice in the next section.

\item {\it Chiral extrapolation}

From the study of the chiral perturbation theory, there appears 
a non-analyticity in the chiral extrapolation at the 
quenched level near the chiral limit as $m_\pi < 400$ MeV \cite{Lei20}.
We have, however, taken the naive linear extrapolation for both the
positive- and negative-parity baryons,
because the results for the baryon masses seem to
be well described with the linear relation in the quark-mass region 
corresponding to $m_\pi > 600$ MeV in the present calculation.
In addition, the behavior of the negative-parity baryon masses
near the chiral limit is less known.
Therefore, it is difficult to estimate the non-analyticity in the chiral
extrapolation from the present results.
The quantitative estimate of its effect on the negative-parity baryon
masses is a future problem with high precision data with small quark masses.

\item {\it Quenching effects}

There is about 10\% uncertainty for the ground-state hadron spectra 
coming from the quenching effect.
Also for the negative-parity baryons, there should appear such a
quenching effect.
In addition, there may appear nontrivial excitation effect of the
$\eta'$ meson in quenched QCD, 
where $\eta'$ degenerates with the other Nambu-Goldstone bosons  
in the chiral limit due to the ignorance of the fermionic determinant.
Such an effect from $\eta'$ is reported to appear 
near the chiral limit as $m_\pi < 300$MeV~\cite{DL03}.

\end{itemize}

\section{Discussion}
 \label{sec:discussion}

Our numerical results for the hadron spectra are summarized in
Fig. \ref{fig:spectrum1} and Table \ref{tab:spectrum_phys2}.
The masses of the negative-parity baryons are found to be heavier than
those of the corresponding positive-parity sectors, as expected.
The flavor-singlet baryon is, however, an exception:
the positive-parity baryon is much heavier than the negative-parity one.
This tendency seems consistent with the 3Q state in the quark model,
in which the flavor-singlet positive-parity baryon belongs to 
the 70-dimensional representation of the $SU(6)$ symmetry with 
the principal quantum number $N=2$.
This multiplet is in general heavier than that belonging to the
negative-parity baryons.
QCD sum rule analysis~\cite{JO96} and the other recent lattice
calculation~\cite{Mel02} also predict the mass of the flavor-singlet 
negative-parity baryon lighter than that of the positive one.

\begin{figure*}
\includegraphics[width=7cm]{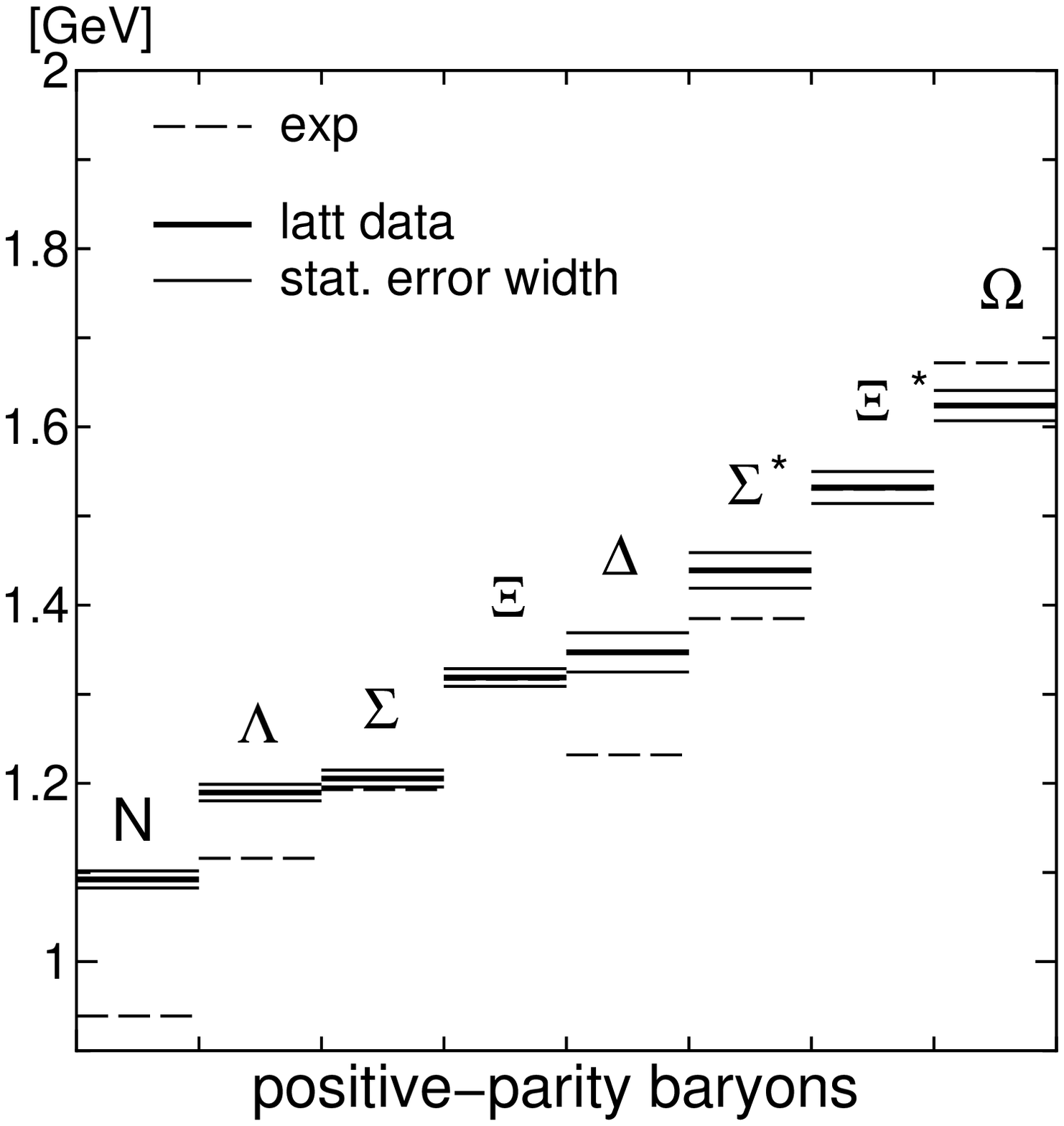}
\includegraphics[width=7.45cm]{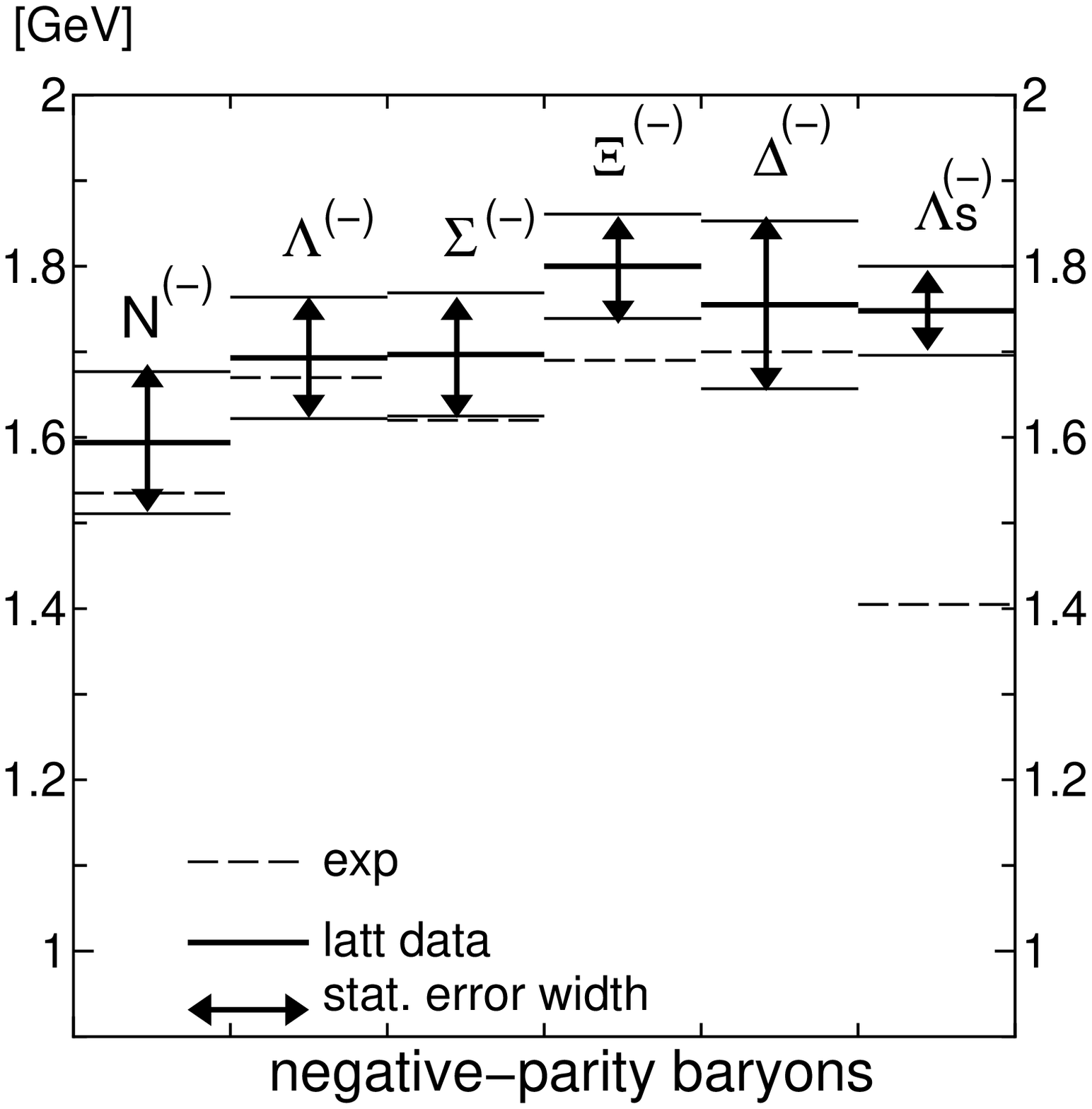}
\caption{
Various baryon masses obtained from the $\beta=6.10$ lattice.
For the negative-parity baryons, the experimental values 
of $N(1535), \Lambda(1670), \Sigma(1620), \Xi(1690), \Delta(1700)$ and 
$\Lambda(1405)$ are added.}
\label{fig:spectrum4}
\end{figure*}

In order to compare our lattice results with experimental values,
various baryon masses at $\beta=6.10$ together with the experimental
values are shown in Fig.~\ref{fig:spectrum4}.
For the positive-parity baryons, the nucleon and the delta masses are somewhat
higher than the experimental ones.
Note again that quenched QCD exhibits the non-analytic behavior in the 
chiral extrapolation on the nucleon near the chiral limit of
$m_\pi < 400$ MeV~\cite{Lei20}.
In comparison with the naive linear extrapolation, 
this effect lowers the nucleon mass in the chiral limit, 
although we have not taken into account the non-analytic behavior 
because of the
absence of its signal in our relatively heavy quark-mass region of 
$m_\pi > 600$MeV.
The other positive-parity baryons with strangeness reproduce the
experimentally observed masses within 10\% deviations.
The better reproduction of strange baryon masses may be natural 
because of the following reason. 
For the strange baryon, the ambiguity from the chiral extrapolation would be 
less than that of the nucleon and the delta, because the strange quark 
is relatively massive and the non-analytic behavior arises only 
from up and down quark mass region.

As for the negative-parity baryons, most of the present 
lattice results comparatively well reproduce the experimental spectra 
as shown in Fig.~\ref{fig:spectrum4}, 
in spite of a relatively large statistical error.
However, the flavor-singlet negative-parity baryon is exceptional, 
and its calculated mass of about 1.7GeV 
is much heavier than the experimentally observed $\Lambda(1405)$ with 
the difference of more than 300 MeV. 
The difference between the lattice result of 1.7GeV 
and the experimental value of the $\Lambda(1405)$ is, however, 
the largest in all the hadrons in consideration.
Even taking the quenching effect into account, this discrepancy 
seems too large.
(Note again that the flavor-singlet baryon has one strange quark, and 
the ambiguity from the chiral extrapolation is expected to be 
less than that of the nucleon and the delta.)

If the $\Lambda(1405)$ could be well described as a three 
valence quark system, 
its mass would be reproduced with the simple three-quark operator 
in quenched QCD. 
However, the light mass of the $\Lambda(1405)$ is not reproduced 
in the present simulation. 
Therefore, the present lattice QCD result physically indicates that 
the experimentally observed $\Lambda(1405)$ cannot be described 
with a simple three valence quark picture, that is, 
the overlap of the $\Lambda(1405)$ with the 3Q component is rather small.
This seems to support other possible pictures for the $\Lambda(1405)$
such as the penta-quark state or the $N\bar{K}$ molecule.
In this sense, lattice QCD simulations with the penta-quark operator 
would be meaningful to elucidate the nature of the $\Lambda(1405)$.

Here, we add several comments and cautions in quenched QCD. 
First, there is possible mixing between 3Q and 5Q states 
through the ``Z-graph" even in the quenched approximation~\cite{Liu99}.
Therefore, to be strict, the separability into 3Q and 5Q states does not hold 
even in the quenched approximation, although the mixing between the two states 
is rather suppressed.
Nevertheless,  the conclusion of the non-3Q picture for the $\Lambda(1405)$ is 
still plausible, because, if the $\Lambda(1405)$ is described as the 3Q state, 
its mass is to be reproduced with the 3Q operator in quenched QCD.
Second, for the definite conclusion, we have to pay attention to the 
non-analytic behavior in the flavor-singlet negative-parity baryon 
near the chiral limit~\cite{Lee02}, in spite of 
the naive expectation of its smaller effect for strange baryons.
Third, according to the neglect of the fermionic determinant,  
$\eta'$ becomes unphysically ``light" as the Nambu-Goldstone particle 
in quenched QCD, and the non-unitary behavior appears in the baryon correlator 
due to the ``light" $\eta'$ excitation near the chiral limit 
below $m_\pi \sim 250$ MeV~\cite{DL03}, 
where there appears an unphysical ``decay'' process of the
negative-parity baryon into an $\eta'$-N state.
Although these non-analytic and the non-unitary behaviors are not 
observed in the present simulation 
with relatively heavy quark masses as $m_\pi >600$MeV, 
these effects should be taken into account for the simulation near 
the chiral limit.

We now focus on other negative-parity baryons.
The mass ratio between the positive- and the negative-parity baryons
is shown in Table \ref{tab:spectrum_phys4}.
For both the octet and the decuplet baryons, 
the relative mass difference between the parity partners becomes smaller,  
as the averaged quark mass increases by the inclusion of the strange quark. 
This tendency is experimentally observed for the octet baryons.
(The empirical identification of negative-parity decuplet baryons 
is not established.)
This behavior is also reported in another lattice QCD analysis
by the domain wall fermion~\cite{SBO01}.
From Fig.~\ref{fig:spectrum4} and Table \ref{tab:spectrum_phys4},
we find that the lattice results of the flavor octet and decuplet baryons
are all close to the observed lowest-lying negative-parity baryons,
the $N(1535)$, $\Lambda(1670)$, $\Sigma(1620)$ and $\Delta(1700)$,
in spite of the relatively large statistical error.
The $\Sigma(1620)$, which is experimentally confirmed as the negative-parity 
strange baryon with $J^P=1/2^-$~\cite{PDG02}, is consistent with the
parity partner of the $\Sigma$ baryon.
The parity partner of the $\Xi$ baryon is expected to be the $\Xi(1690)$ from
our calculation, although the spin-parity of the $\Xi(1690)$ is not yet
confirmed experimentally.
Recently it has been proposed based on the chiral unitary approach
\cite{Ben02} that the $\Xi(1620)$ has the negative-parity,
although its experimental status is still one-star (evidence of
existence is poor).
It is, however, difficult to distinguish between the $\Xi(1620)$ and
the $\Xi(1690)$ from our result due to the statistical errors.
For the decuplet baryons,
we can regard the parity partner of the $\Delta(1232)$ as the $\Delta(1700)$, 
although the experimental data is poor.
The positive-parity flavor-singlet baryon is found to be much heavier 
than the negative-parity decuplet, and hence 
its investigation seems much difficult both theoretically and experimentally.

\begin{table*}
\caption{
The ratio of negative and positive-parity baryon masses.
In the last line, the ratio of flavor-singlet negative-parity and 
octet positive-parity 
baryon masses is also listed. Physical values of the negative-parity
baryons are taken to be $N(1535)$, $\Sigma(1620)$,
$\Xi(1690)$, $\Lambda_{\rm oct}(1670)$, $\Lambda_{\rm sing}(1405)$
and $\Delta(1700)$.
(*)Note that the spin-parity of the $\Xi(1690)$ is not yet confirmed.
}
\begin{ruledtabular}
\begin{tabular}{ccccc}
   & $\beta=5.75$ & $\beta=5.95$ & $\beta=6.10$ &
     physical value \\
\hline
$N^{(-)}/N^{(+)}$            &
                1.516(70)&1.484(54)&1.463(51)& 1.635 \\
$\Sigma^{(-)}/\Sigma^{(+)}$       &
                1.456(54)&1.423(40)&1.410(40)& 1.358 \\
$ \Xi^{(-)}/\Xi^{+}$          &
                1.405(41)&1.373(30)&1.366(31)& 1.282$^*$ \\
\hline
$\Lambda_{\rm oct}^{(-)}/\Lambda_{\rm oct}^{(+)}$&  
                  1.479(54)&1.440(40)&1.417(40)& 1.496 \\
$\Lambda_{\rm sing}^{(-)}/ \Lambda_{\rm sing}^{(+)}$ &
                  0.757(22)&0.718(26)&0.802(32)&  \\
\hline
$\Delta^{(-)}/\Delta^{(+)}$       & 
                 1.322(64)& 1.399(55)& 1.339(61)& 1.380 \\
$\Sigma^*(-)/\Sigma^*(+)$     & 
                 1.299(52)& 1.358(43)& 1.312(49)&  \\
$\Xi^*(-)/\Xi^*(+)$        & 
                 1.280(41)& 1.321(33)& 1.288(38)&  \\
$\Omega(-)/\Omega(+)$       & 
                 1.263(31)& 1.290(24)& 1.266(29)&  \\
\hline
$m_{\Lambda_{\rm sing}^{(-)}}/m_{\Lambda_{\rm oct}^{(+)}}$
               & 
                 1.432(23)&1.392(42)& 1.437(34)& 1.259 \\
\end{tabular}
\end{ruledtabular}
\label{tab:spectrum_phys4}
\end{table*}

Finally, we comment on recent lattice studies on the negative-parity
baryons.
Sasaki {\it et al}. investigated the negative-parity non-strange baryon 
$N^{(-)}$, the parity partner of the nucleon $N^{(+)}$, with 
the domain wall fermion~\cite{SBO01}.
Their lattice is $16^3\times 32$ at $\beta=6.0$ ($a^{-1}=1.9$ GeV) 
and the result is $N^{(-)}/N^{(+)} \sim 1.45$, which is consistent with ours.
G\"ockler {\it et al}. studied the negative-parity non-strange
baryon $N^{(+)}$ using
the $O(a)$ improved Wilson quark on the isotropic lattices with the size of 
$16^3\times32$ and $32^3\times64$~\cite{Goc01}.
They obtained the similar result, $N^{(-)}/N^{(+)}=1.50(3)$.
Melnitchouk {\it el al}. also studied the negative-parity baryons 
using the $O(a)$ improved Wilson quark on the isotropic lattice,
$16^3\times32 (a=0.125$ fm)~\cite{Mel02}.
Since they did not carry out the chiral extrapolation, 
we do not compare the results quantitatively, but the qualitative
behavior is similar to us.
They also investigated the flavor-singlet baryons.
Instead of the flavor-singlet interpolating field, they used the
``common'' interpolating field which is the common part of the
interpolating fields for the
octet $\Lambda$ hyperon and the singlet baryon.
The result is much heavier than the experimental value of the $\Lambda(1405)$
even for such an field.
Lee {\it et al}. investigated the excited state baryons with the overlap
fermion with the lattice $16^3\times28$~\cite{Lee02}.
They employed the constrained curve fitting method for the mass fitting
and obtained the baryon masses lower than those from the conventional
fitting method.
Thus, their results seem to be lower than ours and the other's, while
they did not carry out the chiral extrapolation.
Dynamical quark simulation of excited-state baryons is also in 
progress~\cite{UK02}.
As for the negative-parity nucleon, their present result is consistent 
with the quenched result within statistical errors.

\section{Summary and Concluding Remarks}
  \label{sec:conclusion}

We have studied the mass spectra of the negative-parity baryons
and the flavor-singlet baryons in quenched anisotropic lattice QCD.
We have used three lattices of almost the same physical spatial volume of 
about $(2 {\rm fm})^3$ with the spatial cutoffs $a_{\sigma}^{-1}=1$--2 GeV and
the renormalized anisotropy $\xi=a_\sigma/a_\tau=4$.
We have adopted the standard Wilson plaquette gauge action and the $O(a)$
improved Wilson quark action at the tadpole-improved 
tree-level~\cite{Aniso01b}.
The positive- and negative-parity baryon masses are extracted 
with the parity projection 
from the same baryon correlators based on the three valence quark picture.

For the flavor octet and decuplet negative-parity baryons, 
the calculated masses are close to the experimental values of 
corresponding lowest-lying negative-parity baryons.
For several negative-parity baryons, our lattice data have suggested 
some predictions. 
For instance, $\Delta(1700)$ can be regarded as the parity partner 
of $\Delta(1232)$, and 
the $\Xi(1690)$ would be the parity partner of the $\Xi$ baryon, 
although the spin-parity of the $\Xi(1690)$ is not yet confirmed 
experimentally.

As for the flavor-singlet negative-parity baryon, 
such a three-quark state has been found to lie around 1.7GeV, 
and has been much heavier than the $\Lambda(1405)$. 
Even considering the systematic errors, this difference of about 
300 MeV seems too large.
If the $\Lambda(1405)$ is described as a three valence-quark state, 
its mass would be reproduced in the present simulation. 
In fact, the present lattice result which cannot reproduce the $\Lambda(1405)$ 
physically implies that the $\Lambda(1405)$ is not described as the 
simple three quark picture, 
i.e., the overlap of the $\Lambda(1405)$ with the three-quark state is 
rather small.
This seems to support an interesting picture of the penta-quark state 
or the $N\bar{K}$ molecule for the $\Lambda(1405)$.
For more definite understanding of the $\Lambda(1405)$,
it would be desired to perform lattice QCD simulations in terms of 
the $N\bar{K}$ molecule or the penta-quark state.
Such a study is interesting even at the quenched level, where
dynamical quark loop effect is absent and then
the quark-level constitution of hadrons is clearer. 
As for the positive-parity flavor-singlet baryon, its calculated 
result is found to be much heavier 
than the negative-parity one, as is consistent with 
the quark models~\cite{IK79,CI86} 
and the QCD sum rule analysis~\cite{JO96}.

Very recently, LEPS Collaboration has experimentally observed 
the $\Theta^{+}$ (or $Z^+$) baryon with $S=+1$~\cite{N03}, 
which requires at least ``five valence quarks" as ${\rm uudd}\bar {\rm s}$
and is physically identified as a ``penta-quark system".
The comparison between the $\Theta^{+}$ baryon and the $\Lambda(1405)$ 
may be useful to investigate the features of the penta-quark system.
It is also interesting to investigate this type of a penta-quark
system using lattice QCD simulation.

From aspect of the chiral symmetry, the parity partner should be degenerate 
if the symmetry is restored at finite temperature and/or density.
It is interesting to see how the mass difference between the parity partners
changes at finite temperature on the lattice.
Several works on it are already reported for the screening mass of the
nucleon~\cite{DK87}
and they favor the parity degeneracy at the chiral phase transition.
Recently based on the chiral effective theory such as the linear sigma
model~\cite{JI00} and the chiral perturbation theory~\cite{NE97},
two different assignments for the negative-parity baryons have
been proposed: under the chiral transformation,
the negative-parity baryon transforms in the same way as the 
positive-parity one in one scheme and in the opposite way in the other.
These two assignments behave differently toward the chiral restoration.
Therefore, it is interesting to study them from the quark degrees of freedom
such as in lattice QCD at finite temperature.

\section*{Acknowledgments}

Y.N. thanks S. Sasaki, T. Blum and S. Ohta for useful discussions and comments.
H.M. thanks T. Onogi and T. Umeda for useful discussions.
The simulation was done on
NEC SX-5 at Research Center for Nuclear Physics, Osaka University and
Hitachi SR8000 at KEK (High Energy Accelerator Research Organization).
H.S. is supported in part by Grant for Scientific Reserch
(No.12640274) from Ministry of Education, 
Culture, Science and Technology, Japan.   
H.M. is supported by Japan Society for the Promotion of Science
for Young Scientists.
Y.N. was supported by the center-of-excellence (COE) program
at YITP, Kyoto University in most stage of this work and
thanks RIKEN, Brookhaven National Laboratory and the U.S. 
Department of Energy for providing the facilities essential for the
completion of this work.

\end{document}